\documentclass[11pt]{elsarticle}

\usepackage{amsmath}
\usepackage{soul}
\usepackage{xcolor}
\usepackage{hyperref}

\usepackage{geometry}
\usepackage[english]{babel}
\usepackage{microtype}

\newcommand{\blue}[1]{{\color{black}#1}}

\title{ACCES: Non-Invasive Simulation Calibration via Optimisation using Evolutionary Algorithms and Metaprogramming} 

\author[inst2]{Andrei-Leonard Nicușan}
\author[inst2]{Dominik Werner}
\author[inst1,inst2]{Jack A. Sykes}
\author[inst2]{Jonathan Seville}
\author[inst1]{Tzany Kokalova}
\author[inst2]{Christopher R. K. Windows-Yule}

\affiliation[inst1]{organization={School of Physics and Astronomy},
            addressline={University of Birmingham}, 
            city={Birmingham},
            postcode={B15 2TT}, 
            country={UK}}

\affiliation[inst2]{organization={School of Chemical Engineering},
            addressline={University of Birmingham}, 
            city={Birmingham},
            postcode={B15 2TT}, 
            country={UK}}

\begin{document}

\maketitle 

\\[1em]

\noindent \textbf{Abstract}: Calibration is a vital step in the development of rigorous digital models of diverse physical and chemical processes, yet one which is highly time- and labour-intensive. In this paper, we introduce a novel tool, Autonomous Calibration and Characterisation via Evolutionary Simulation (ACCES), which uses evolutionary optimisation to efficiently, autonomously determine the calibration parameters of a numerical simulation using only simple experimental measurements. 
Modelling calibration as an optimisation problem, it can use any combination of sensibly-chosen experimental measurements as calibration targets for virtually any numerical modelling technique, e.g. computational fluid dynamics (CFD), finite element analysis (FEA), discrete element method (DEM). We provide a case study demonstrating the application of the method to DEM modelling and comprehensively validate the calibrated DEM particle properties using positron emission particle tracking (PEPT) experiments in conjunction with DEM digital models of systems spanning orders-of-magnitude differences in scale and covering both collisional and frictional operating regimes, showing excellent agreement. Key results show that i) even simple, well-chosen targets can achieve highly-accurate calibration, ii) though the calibration process only used images corresponding exclusively to a free-flowing surface shape, it was able to reproduce completely different quantities as well, such as velocities and occupancies, with quantitative precision, and thus iii) DEM can quantitatively predict complex particulate dynamics even outside the calibration system. ACCES is a portable, tested, documented and officially-registered Python package available under the GNU General Public License v3.0.

\newgeometry{margin=1.5cm}
\twocolumn

\section{Introduction}
\label{sec:introduction}

An important problem in the field of numerical modelling is \textit{simulation calibration}: given a simulation of a physical system, what microscopic model parameters are required to reproduce experimental, macroscopic measurements? \cite{windows2022calibration, kennedy2001bayesian}. {For consistency, we define ``microscopic model parameters'' as local, intrinsic properties used by a computer model which are non-trivially detached from ``macroscopic'', bulk, or global phenomena exhibited by the model (e.g. the effect of Non-Newtonian viscosity, a micro parameter, on cavern formation in stirred tanks, a macro observation)}. A complicating factor in the \textit{calibration} of such models is that oftentimes a single simulation can take from hours to weeks to run, and may require expensive, supercomputing resources. As such simulations often have significant numbers of free parameters to characterise, it is intractable to explore the complete parameter space via a conventional, grid-based design of experiments; in other words, the full effects of all \textit{micro} parameters on the \textit{macro} measurements are unquantifiable at best, or completely unknown at worst \cite{plumlee2019computer}. However, without correctly calibrating a numerical model of a system, its results not only cannot be trusted, but may be completely unphysical -- or, worse, incorrect yet insidiously sensible-looking -- which can harmfully affect future design choices. In a more positive light, once the physics have been accurately characterised, the system can be broadly investigated and optimised without requiring new experimental measurements or recalibration, even at scales that would be unfeasible to construct for experimental imaging \cite{WINDOWSYULE2024}.

The challenge of model calibration is particularly pronounced in the case of discrete element method (DEM) simulations, which allow the detailed numerical modelling of the dynamics and mechanics of particulate systems \cite{luding2008introduction}. DEM simulations are widely -- and ever-increasingly -- used in both academia and industry to model a diverse range of processes, from earthquakes and avalanches \cite{tang2009tsaoling} to pharmaceutical secondary manufacture \cite{ketterhagen2009process} and even the movements of human beings \cite{smith2009modelling}. The popularity of DEM stems largely from the fact that it is capable of providing more, and more detailed, information than any contemporary experimental technique \cite{rosato2020segregation}, and -- if rigorously calibrated \cite{angus2020calibrating,lumay2012measuring,allen2013particle,mangwandi2007coefficient} -- can do so with quantitative accuracy \cite{windows2022calibration,windows2016numerical}. It is this `if', however, that provides biggest drawback of DEM: without careful calibration of the various microscopic parameters used in DEM models, their outputs simply cannot be trusted \cite{windows2016numerical}. Typical calibration experiments, requiring the characterisation of diverse particle properties including size, density, Young's modulus, restitution coefficients (both normal and tangential, inter-particle and particle-wall) and frictional coefficients (rolling, sliding and torsional, inter-particle and particle-wall), are inherently time-consuming and require numerous different experimental apparatus, each of which requires expert operation to yield accurate values\cite{angus2020calibrating,lumay2012measuring,allen2013particle,mangwandi2007coefficient,rosato2020segregation}. 

In this paper, we detail a novel methodology, \textit{Autonomous Characterisation and Calibration using Evolutionary Simulation} (ACCES), which couples arbitrary simulation software to an evolutionary algorithm which -- based on one or more simple reference experiments -- is able to fully-autonomously `learn' the most suitable set of simulation parameters to accurately model the physical system of interest. ACCES models calibration as an optimisation problem, and employs a unique metaprogramming-based software architecture which uses entire simulation \textit{scripts}, as opposed to simple mathematical \textit{functions} (as employed by virtually all contemporary optimisation packages). ACCES can thus work with arbitrarily complex simulation and post-processing workflows. It can calibrate simulation parameters of interest based on any experimentally-measurable property, from any sensibly-chosen experimental apparatus, using any viable imaging method, from PIV \cite{adrian1991particle} to positron emission particle tracking \cite{windows2020positron}.

\subsection{Alternative Calibration Approaches}
\label{sec:alternative}



The simulation, or prediction, of physical phenomena can be divided into two broad categories: data-driven \textit{statistical models}, which assume model exactness and augment it with additional terms to account for inaccuracy and noise, and more refined \textit{computer models} which exhibit reasonable physical trends, even if the model itself is slightly inexact \cite{plumlee2019computer, kennedy2001bayesian}. Statistical models, in general, are more straightforward to build and are mainly concerned with prediction over the given data-fitting range, but require an expansive set of samples to fit over; computer models lie at the opposite end of the spectrum: they require more sophisticated modelling of the underlying physics to provide predictive properties \textit{outside} the initial experimental samples \cite{campbell2006statistical, higdon2004combining}. This paper is concerned with the calibration of the latter, with a focus on novel non-invasive software architectures for calibration via optimisation, and an example of its application to discrete element method simulations of granular systems.

The calibration of computer models can further be divided into two: direct measurements of physical \textit{micro} properties, and indirect calibration against \textit{macro} measurements \cite{windows2022calibration}. The former is fairly limited to fewer, relatively simple properties: for example, the viscosity of Newtonian fluids can be directly measured and implemented in direct numerical simulations (DNS) solving the constitutive Navier-Stokes equations at all time- and length-scales, but this is limited to relatively small systems due to the extreme computational expense of the technique \cite{moin1998direct}; using more complex rheological models, or equations of state for compressible fluids, makes the ``inexact-ness'' of the computer model more prominent and shifts the scale towards requiring indirect, macro-scale calibration -- the precise point at which this transition happens depends on the complexity and scales of the system being modelled, and there is an extensive body of literature exploring both \cite{alfonsi2011direct, pusok2018effect, guillou2007effect, li2010direct, nikolaou2015direct}. However, fluid simulations are perhaps different to most other computer modelling fields, in that the constitutive model is well-known; for more inexact computer models, such as discrete element method simulations of granular dynamics, where there is no widely-accepted continuum model, direct calibration has far less, and in some cases no, applicability \cite{windows2022calibration}.

This work thus focuses on the indirect calibration of complex computer models against experimental measurements. In the following paragraphs we discuss alternative calibration approaches used in some fields employing simulations of physical systems, and their advantages and disadvantages.

\subsubsection{Grid Calibration}

Perhaps the most straightforward calibration method -- called grid-based, manual, or brute-force -- employs a fixed set of input parameter values, typically as equidistant points spanning their possible ranges, which are all evaluated, i.e. a simulation is run for every parameter combination, with the one producing the lowest discrepancy to the experimental measurement being taken as the calibrated values. This approach was used, for example, for calibrating finite element structural simulations of concrete \cite{milligan2020finite, xu2020calibration}, finite difference simulations of surface water-groundwater interaction \cite{jafari2021fully}, as well as discrete element method simulations of powders in a one-factor-at-a-time approach \cite{simons2015ring, wensrich2012rolling}. While simple to implement, this approach has two significant disadvantages: first, it is strongly affected by ``the curse of dimensionality'', wherein more calibration parameters result in an exponential increase in the number of samples required; e.g. while running simulations on a $10 \times 10$ grid for two input parameters is reasonable, for six inputs (which are not at all many for e.g. DEM simulations \cite{coetzee2017calibration}) the equivalent approach would require one million simulations, which, even for medium-scale computer models would result in completely intractable computational expense. Second, results are highly dependent on the grid spacing -- in the previous example, 10 increments between the bounds of a parameter may well be too few and miss the real parameters of best fit; a prime example is found in Section \ref{sec:converging} of this paper, where two parameters with ranges between $(0, 1)$ are found to be halfway between two such hypothetical increments. Careful placement of sampling simulations thus becomes necessary for more calibration parameters and complex computer models.

\subsubsection{Surrogate Model Calibration}


The next broad category of calibration techniques involve the use of a \textit{surrogate model}, or metamodel, as a way to convert discrete sampling points into a continuous function which is much faster to evaluate than the complex computer models; this surrogate model is then iteratively interrogated, typically via some form of root-finding to calibrate the metamodel, and thus, indirectly, the simulation \cite{owen2017comparison}. Perhaps the oldest approach in this category still employed is that of classical design of experiments (DoE) with response-surface methodologies; in practice, this often involves fitting second-order multivariate polynomials to sampling points evaluated over a given design - for example, a simplified RANS CFD model of urban airflow was calibrated in this manner using NLPQLP optimisation as the root-finder on the fitted response surface \cite{shirzadi2020rans}. Simpler DoE approaches have been similarly employed to calibrate four parameters in a CFD simulation of diesel-fuel spray atomisation \cite{brulatout2016calibration}, or CFD simulations of drag over automotive shapes \cite{williams1994calibration}. DoE-based calibration has been used for DEM simulations of powders, for example in calibrating 7 microscopic parameters at 3 levels (i.e. increments between input bounds) against a single uniaxial test \cite{yoon2007application}; however, recent research revealed the extremely non-linear nature of the effects of DEM contact parameters and, put bluntly, the failure of quasi-static measurements alone to calibrate them due to the extreme underdetermination of the calibration problem, especially upon subsequent validation in dynamic systems \cite{windows2022calibration}. This highlights the main problem with classical DoE-based calibration and optimisation: when applied to moderately complex computer models, the low-order multivariate polynomials fitted simply cannot capture the much more complex response surfaces of the former (which are typically much more nonlinear and noisy) - and thus the optima found are simply artefacts of the incorrect response surface fitted.

More sophisticated sampling methods and surrogate models have been used to higher degrees of success: for example, neural networks have been an attractive option for surrogates due to their ability to be universal function approximators, given enough layers and extensive training data \cite{sonoda2017neural}. Neural networks have been used as surrogates for CFD modelling of data centres \cite{wang2020kalibre}, stress distributions and structural reliability in finite element simulations \cite{liang2018deep, lieu2022adaptive}. In discrete element simulations of powders, neural networks have similarly been used for calibration, with the ``extensive training data'' requirement becoming apparent: in one study, 546 shear cell measurements were needed to train a surrogate for 4 input parameters \cite{benvenuti2016identification}; in another, 300 samples placed using Latin Hypercube Sampling were used to train a neural network against two calibration targets \cite{ye2019calibration}. Another popular surrogate modelling approach includes the use of kriging, or general Gaussian processes to interpolate between samples; an advantage of these techniques is that they yield direct uncertainty measures of the predictions in the form of variances \cite{kleijnen2017regression}. Examples include the calibration of cable-stayed bridge simulations using finite element analysis \cite{zhang2014calibration} and calibration of combined finite-discrete element method simulations of deformable particles against compressive and tensile strength measurements \cite{lei2024efficient} They have been used in DEM to interpolate between 72 samples placed based on a 3-level design of experiments to predict static angles of repose; following the discussion in the previous paragraph, it is unlikely that they are enough to provide closure for 6 calibration parameters \cite{rackl2017methodical}.

Perhaps the most popular and actively-worked-on technique in modern calibration is based on Bayesian optimisation, which provides arguably the most consistent mathematical description of model uncertainty with respect to its input parameters against experimental observations \cite{kennedy2001bayesian, plumlee2019computer}. It has been used in a wide variety of contexts, including the calibration of CFD-based fluid flow, turbulence, combustion and multiphase modelling \cite{guillas2014bayesian, duan2021fixed, wu2016bayesian, liu2021uncertainty, liu2019validation}, FEM-based laser-melting, structural analysis and soft-tissue simulation \cite{kusano2021novel, ezzat2020model, ebrahimian2017nonlinear, ramancha2020bayesian, hizal2020two, madireddy2016bayesian}. Its application in DEM seems to suffer from the same disadvantages as the previous surrogate-based approaches: for example, 2000 DEM simulations of a quasi-static triaxial test were used for calibrating 5 parameters \cite{cheng2018probabilistic}. While a commonly-cited disadvantage of the technique is the computational expense of the surrogate model, for example in generating new samples to evaluate, this cost is typically orders of magnitude smaller than that of even moderately-complex computer models, and therefore insignificant. While mathematically sound, the greatest disadvantages in its application to complex computer models is the need for expert knowledge to build a Baysian-based surrogate model for effectively every class of simulation; the choice of priors, the key building blocks of the probabilistic formulation of calibration, is strongly non-trivial and significantly affects the uncertainty quantification results, typically requiring extensive prior art, expert knowledge or large sampling sets to work effectively \cite{kennedy2001bayesian}.

\blue{
Note that in this case the ``expert knowledge'' required by e.g. Bayesian optimisers and Gaussian Processes refers to the correct mathematical formulation that the underlying optimisation problem should follow - e.g. what combination of kernels reproduce a given granular system response that is to be calibrated; it is unknown if the same kernels should reproduce all granular systems, or even different regimes within one - or the metaparameters that should carefully be tuned for specific problems. This is distinct to the knowledge required to choose a sensible set of calibration instruments for the given parameters that are to be calibrated, which would be required for any calibration approach, irrespective of the calibrator itself; for example, the methodology described in Section \ref{sec:methodology} could well be used with a different choice of calibrator, including a surrogate model.
}

To summarise, virtually all surrogate approaches exhibit similar disadvantages: i) the entire parameter space must be sampled in order to provide prediction for new operating states, significantly increasing their computational expense during the training stage, ii) new operating states must be experimentally-imaged in precisely the same manner in order to have valid predictions, which can be difficult depending on the computer model; for example, discrete element method simulations of the popular dynamic angle of repose powder characterisation measurement is sensitive to the number of particles included in the system, which is very difficult to control experimentally, iii) sampling is often unguided, done in one batch, with effectively equal computational effort spent over unfeasible parameters, as close to the true calibrated values, iv) sampling may easily miss the global optimum, and instead we rely on the surrogate model to approximate the optimal region effectively; however, surrogate model errors increase the further predictions are from directly-sampled points, and v) implementing a simulation-in-the-loop architecture for automated, on-demand parameter sampling requires invasive rewriting of simulations as a function (in the programming sense) to be evaluated.

\subsubsection{ACCES Mitigations and Unique Aspects}

ACCES solves the above issues in the following ways:

\begin{enumerate}
    \item We combine the advantages of batch-based sampling (i.e. parallel execution efficiency) and optimised sample placement: unlike Bayesian sampling of one parameter combination at a time, ACCES generates a family of samples in epochs, with each epoch focusing on zooming onto the global optimum.
    \item Thus, as sampling is guided by an evolutionary optimiser, we do not have to finely-sample all regions of the parameter space; instead, we focus on discovering the globally-optimum calibrated parameters in as few samples as possible.
    \item Rather than sampling around the whole parameter space to build a surrogate model which becomes more inaccurate away from the directly-sampled points, we only use direct sampling, ensuring exact correspondence between ``prediction'' and computer model output.
    \item Therefore, calibration is not sensitive to different system operating states to previous runs -- calibration is conducted ``afresh'' for new states and runs.
    \item While evolutionary algorithms excel in derivative-free, blackbox optimisation, they are notorious for requiring great numbers of function evaluations -- which, in this case, correspond to entire computer model evaluations. We use state-of-the-art CMA-ES strategies -- augmented in this work for calibration purposes (see Section \ref{sec:ACCES}) -- which use fewer than 100 evaluations per free parameter, being competitive even with classical gradient-based optimisers; e.g. in Section \ref{sec:converging}, 4 parameters are calibrated in 144 samples.
    \item Unlike Bayesian-based calibration, the ``calibrator does not need to be calibrated'' -- i.e. there are effectively no algorithm metaparameters that must be tuned for the specific system of interest; as detailed in Section \ref{sec:ACCES}, in practice no ACCES metaparameters affect the overall success of a calibration run, so long as the simulation and experimental measurements are sensibly-specified.
    \item ACCES implements a novel simulation-in-the-loop architecture that works with virtually any simulation engine; \blue{it is non-invasive in that it does not require rewriting simulations as functions, or reformulating them in a different programming format.}
\end{enumerate}

The main disadvantage in relation to other approaches is that uncertainty quantification is not directly offered by the calibration algorithm itself. However, uncertainty in Bayesian optimisation is strongly dependent on the choice of prior, which is especially difficult to choose for complex computer models where prior art is limited; if not many samples are available, as is the case for computationally-expensive models, the choice of prior becomes even more influent \cite{kennedy2001bayesian}. A similar trend is found based on the choice of kernels in Gaussian Processes \cite{kleijnen2017regression}. Therefore, it may be argued the \textit{absolute} values yielded by uncertainty quantification in such cases should be carefully assessed, and instead, the \textit{relative} values may be more trustworthy. On the other hand, relative assessment of parameter uncertainty and importance can be offered by modern sensitivity analysis tools, which can be applied directly to the samples tried by ACCES after a calibration run, offering arguably richer information, such as higher-order interactions by global sensitivity indices computation techniques like High-Dimensional Model Representation (HDMR) \cite{li2010global}.

\subsection{Significance of this Work}

We describe the general software architecture in Section \ref{sec:ACCES}, then demonstrate its application by solving the infamously difficult problem of discrete element method (DEM) calibration, coupling ACCES with simulation and post-processing packages written in 4 different programming languages, showing its flexibility. We use simple, binary, optical data acquired from \textit{multiple} rotating drum measurements \cite{lumay2012measuring}, demonstrating that highly accurate calibration can be achieved even with extremely simple inputs.
We prove the identification of the globally-optimum physical parameters -- i.e. correctly back-computing the microscopic particle parameters which reproduce multiple macroscopic rotating drum operating regimes simultaneously -- by validating the calibrated powder properties against experiments (Section \ref{sec:validation}) covering:

\begin{enumerate}
    \item Different rotating operating conditions.
    \item Experimental quantities that were not used for calibration (e.g. velocity).
    \item Completely different systems -- the vibrofluidised bed and tumbling mixer -- of orders-of-magnitude differences in scale, covering frictional and collisional, as well as low- and high-density flows.
\end{enumerate}

We thus also comprehensively prove that correctly-calibrated DEM particle parameters can quantitatively simulate systems operating in completely different regimes and scales to the calibration systems, addressing a significant open question in the field. This has two further key implications for DEM users in academia and industry: 1) DEM, while approximating some physical powder characteristics (e.g. spherical particle shapes, inter-penetration-based contact models), \textit{does} indeed quantitatively simulate the bulk behaviour with predictive capabilities, including outside calibration ranges {(even accounting for the aforementioned inexactness; see discussion in Section \ref{sec:alternative})}, 
and 2) industrial-scale systems can be modelled accurately \textit{and} cost-effectively by calibrating the handled powders in a smaller, faster-to-simulate calibration system. 

While ACCES and its architecture have originally been developed to solve DEM powder characterisation -- which are detailed and extensively tested in the present paper -- the software itself is a general-purpose calibration and optimisation framework, and has been used successfully across different fields, such as for calibrating computational fluid dynamics  \cite{hart2024autonomous} and Monte Carlo \cite{herald2022autonomous, herald2023monte} simulations.

An online, interactive course on general calibration and optimisation using ACCES, discrete element method simulations, and coupling the two is available {\href{https://github.com/uob-positron-imaging-centre/ACCES-GranuDrum-Calibration}{here}}. The library is published to the {\href{https://pypi.org/project/coexist/}{Python Package Index}} and its source code is available under the GNU General Public License v3.0 on {\href{https://github.com/uob-positron-imaging-centre/ACCES-CoExSiST}{GitHub}}.

\section{ACCES: Autonomous Calibration and Characterisation using Evolutionary Simulation}
\label{sec:ACCES}

\begin{figure*}[htbp]
    \centering
    \includegraphics[width=\linewidth]{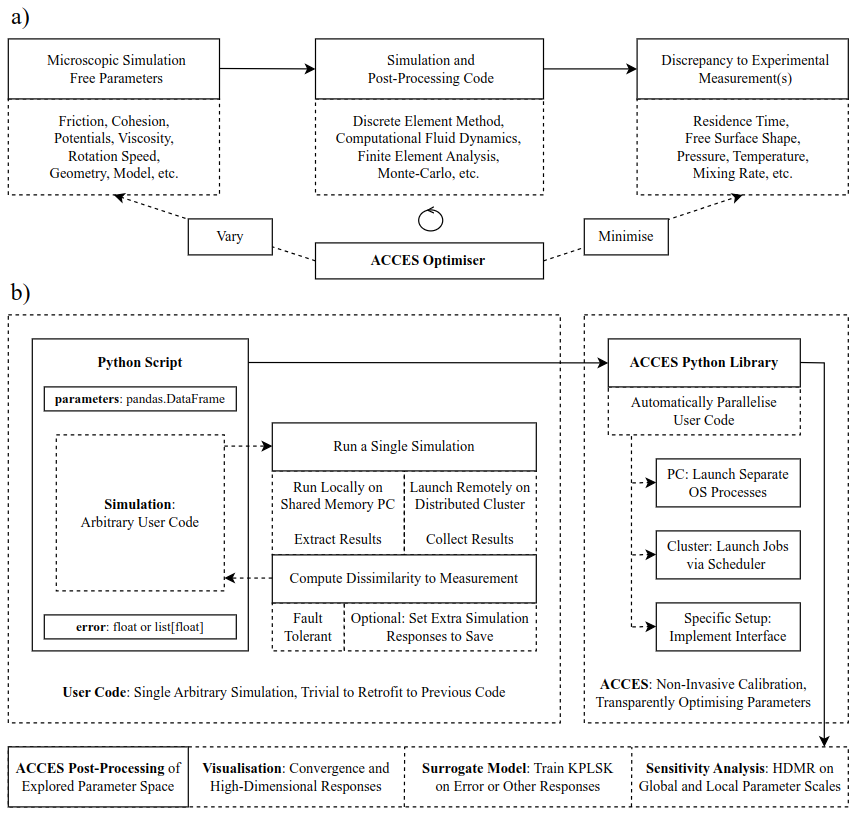}
    \caption{Diagram illustrating the ACCES calibration and optimisation architecture; the high-level steps depicted are implemented in portable pure Python code, gracefully handling OS-specific signals, filesystems and process management. There are no limitations imposed on the user-defined simulation to calibrate, allowing different programming languages to coexist, thread-unsafe libraries to be loaded concurrently or multi-level parallelism to be achieved on remote, distributed computing clusters. {Panel $a)$ illustrates the general idea of modelling calibration as an optimisation problem, and having a ``simulation in the loop''. Panel $b)$ shows a schematic of the ACCES architecture, starting from an arbitrary script, which ACCES automatically parallelises to run and calibrate across local or cluster-based computing architectures.}}
    \label{fig:architecture}
\end{figure*}

In this section we describe the Autonomous Calibration and Characterisation via Evolutionary Simulation -- or ACCES -- framework in its general-purpose form, focusing on the methodology and software architecture. Its concrete application to discrete element method simulations of granular systems and their validation is detailed in the later sections.

\subsection{Calibration as an Optimisation Problem}

We model calibration as an optimisation problem, framing it as the answer to ``what parameter values minimise the discrepancy between simulation and experiment?''. The general idea is illustrated in panel $a)$ of Figure \ref{fig:architecture}. As an optimisation problem, calibration has a few defining characteristics:

\begin{itemize}
    \item The input free parameters may span multiple orders of magnitude, e.g. sliding friction between $\mu_s \in (0, 1)$ and number of particles in the system $N_P \sim \mathcal{O}(1000, >1,000,000)$.
    \item As $N_P$ in the above example, some may be mixed-integer.
    \item A function evaluation corresponds to running one or multiple entire simulations, which may take hours to days to execute.
    \item The outputs of simulations of physical systems, like experimental measurements, are expected to be noisy, non-smooth, and strongly non-linear.
    \item If multiple experimental measurements are to be matched - as is often necessary for good calibration (see Section \ref{sec:methodology}) - it becomes, in essence, a multi-objective problem. However, as correct calibration should yield a single set of parameters, the objectives may be ``scalarised'' into a single target.
\end{itemize}

Thus, it is a very difficult, expensive global optimisation problem, for which the vast majority of optimisation literature -- pertaining to gradient-based optimisation -- is unapplicable.

The correctness of scalarising multiple calibration objective functions into one is more subtle: if 1) the physics of the system being simulated are fully captured (even if some may be approximated), and 2) the calibration measurements correctly cover and show sensitivity to the input parameters, then there should be a unique set of physical parameters yielded by calibration, corresponding to a single global minimum in the calibration objective. Related to 2), if some inputs are confounded -- e.g. often the particle sliding friction can be traded off with increasing rolling friction in DEM \cite{windows2022calibration} -- a suitable set of measurements must be chosen that provide enough ``closure relations'' to fix the inputs to unique, physical parameter values. The converse is also true: if multiple parameter sets yield very similar objective values, or there is a non-dominated Pareto front for the inputs, then the calibration cannot succeed, and instead a set of parameters that reproduce the calibration measurement, but will fail to correctly model different systems (with e.g. different flow states, operating regimes), will be identified. These points will be treated at length in the sections on DEM calibration of powder-handling, a field famous for the difficulty of choosing calibration measurements.

This has positive implications for the choice of optimiser though, as single-objective problems require orders of magnitude fewer function evaluations than multi-objective problems, where typically $\mathcal{O}(10,000-100,000)$ function evaluations are required to identify a Pareto front. As there should not be a Pareto front in the first place in calibration, we can use optimisers requiring $\mathcal{O}(10-1000)$ function evaluations, and hence drastically reduce the number of simulations that need to be run.

\subsection{Software Architecture}

The ACCES algorithm was implemented as part of the open-source {simulation calibration library \textit{Coexist} developed by the authors}, 
written in the Python programming language using an object-oriented design pattern and metaprogramming techniques, allowing for a great degree of flexibility and reusability. The library can naturally accommodate arbitrary simulation engines and complex error functions \textit{without requiring library source-code modifications}. It leverages the ecosystem of foundational Python libraries, sharing features and ensuring compatibility with ubiquitous packages such as \textit{NumPy} and \textit{Pandas} \cite{harris2020array, virtanen2020scipy, bressert2012scipy}. ACCES provides an interface that is easy to use, but powerful enough to automatically parallelise arbitrary user scripts through code inspection and runtime script generation. It can be successfully used from laptop-scale shared-memory machines to multi-node supercomputing clusters.

ACCES is designed to handle a range of issues alongside basic optimisation, including:
\begin{itemize}
    \item Automatically parallelising arbitrary user code using metaprogramming.
    \item Modifying user parameters at runtime to optimiser-predicted trials.
    \item Combining multiple objectives into a single, scalarised target.
    \item Scaling of input parameters with orders of magnitude difference into a consistent phenotype space.
    \item Managing unstable, or crashing simulations.
    \item Saving optimisation progress to return to the last completed epoch in case of node failure.
    \item Scheduling and coordinating the running computational jobs.
    \item Generating visualisations for analysis while calibration / optimisation is running.
\end{itemize}

\subsubsection{The ACCES User Script}

Virtually all optimisation frameworks require users to translate their simulations into standard functions -- within the programming language through which the framework is accessed -- that will be called by the optimisation subroutine. While more manageable from mathematical and software development points of view, this approach is unsuitable for simulation calibration, as simulations generally involve complex scripts set up in multiple different programming or domain-specific languages, calling into optimised compiled code, which is oftentimes run on different machines in a company-wide server environment. Translating a simulation into a standard function format is thus prohibitively difficult. ACCES takes a different approach: instead of using functions to be optimised, it accepts entire \textit{scripts}.

As illustrated in panel $b)$ of Figure \ref{fig:architecture}, a user sets up a standalone simulation script that \textit{can be run on its own} to produce a single function evaluation and corresponding error value; the free parameters to be varied are given in a variable named exactly \textit{parameters}, which is a {Pandas DataFrame} containing parameter names, acceptable bounds and any other columns needed by the user; finally, the error value is stored as a number (for a single calibration measurement) or a list of numbers (for multiple targets) in a variable named exactly \textit{error} which can be defined anywhere in the script. Note that there are no limitations on how exactly \textit{parameters} and \textit{error} are defined, nor what the simulation script actually does - it could simply run a simulation defined inline, call a different program altogether, or authenticate on a server, launch a simulation there and collect results asynchronously. The script-based approach allows straightforward development of calibration runs: if a single simulation works, the entire optimisation can be done without tedious, error-prone format translation steps; moreover, pre-existing simulation scripts can be calibrated by simply adding the free parameters to the \textit{parameters} variable and defining an \textit{error} value to be minimised.

\subsubsection{Metaprogramming / Code Generation}

Once a user simulation script can be run on its own, producing a single ``function'' evaluation and error value(s), it can be calibrated by running the script with different free parameter predictions generated by the optimiser, whose corresponding error values are fed back to the optimiser to generate a new family of predictions and thereby iteratively find the free parameter values which minimise the error. This process of setting up the calibration with different free parameter values is done in two steps: first, the user simulation script is modified to change the free parameter values and save the computed cost to disk, then different parameter value combinations are evaluated in parallel by \textit{scheduling} simulation runs (e.g. different OS processes or distributed cluster jobs).

The original user simulation script code is inspected, extracting the \textit{parameters} definition and changing that section of code from \textit{defining} them to \textit{loading parameters} from an ACCES-defined location. The generated simulation script is run with three command line arguments: a path to the \textit{parameters} saved to disk, a path to save the computed \textit{error} to and another path for an optional \textit{extra} variable containing any additional data a user wants to save for each simulation run (e.g. particle locations). The code inspection is done at the abstract syntax tree (AST) level, the program representation yielded by parsing code into tokens corresponding to each legal Python construct (e.g. variable assignment, function definition). The AST node containing the assignment to a variable named \textit{parameters} is switched to a node loading them from the first command line argument; that section of code is unparsed back into source code representation and injected into the generated simulation script. Finally, code saving the \textit{error} and, if defined, the \textit{extra} variable to disk is appended after the user script. All saving to and loading from disk is done in the ``Pickle'' format which can serialise arbitrary Python objects, including data from compiled extensions such as NumPy arrays if suitable object methods are defined; note that the \textit{parameters} variable is a standard Pandas DataFrame and the \textit{error} variable is a simple floating-point number or list thereof.

\subsubsection{Parallel Simulation Execution}

The generated user script, corresponding to a single function evaluation for a given combination of free parameter values, can be run in parallel, independently from all other script evaluations. Simulations of physical systems are generally extremely computationally-intensive and hence need to make use of parallel computing architectures, from multi-processor shared-memory machines (e.g. laptops, workstations) to distributed clusters. In these cases, the ability to run calibration simulations in parallel becomes essential; this is achieved via a job \textit{scheduling} interface,
which executes the generated script in an arbitrary environment. Two schedulers are offered by default: a local paralleliser and a supercomputing SLURM-based scheduler. The local machine ``scheduler'' is the Python interpreter itself, whose execution is launched in a separate OS Process which is automatically scheduled by the Operating System to run on a local device following the best core-scheduling policy offered by the local kernel. A more complex example of a scheduler interface implementation involves the scheduler submitting jobs to a supercomputing cluster workload manager such as SLURM to execute the generated Python script on a different machine altogether \cite{yoo2003slurm}. This architecture has two main advantages:

\begin{enumerate}
    \item ACCES calibration transparently scales from smaller calibration problems run locally to large-scale supercomputing-level simulations.
    \item The OS process-level separation of simulation execution means that simulation crashes are gracefully handled (and thus unfeasible parameter regions are naturally identified and avoided) and thread-unsafe simulation engines, such as LIGGGHTS, can safely be used concurrently.
\end{enumerate}

\subsubsection{Evolutionary Optimiser}

The core optimiser of the ACCES library is formed by the Covariance Matrix Adaptation Evolutionary Strategy (CMA-ES), a global, derivative-free optimisation algorithm following Darwinian evolution principles, in which generations of parameter combinations are iteratively evaluated and improved in order to minimise an objective function \cite{hansen2006cma}. CMA-ES fits a multivariate Gaussian distribution to the sampled space, where each free parameter corresponds to a dimension; the use of a distribution, rather than a function, is particularly powerful as it allows the function (in this case, simulation) responses to be noisy, non-smooth and discontinuous, as opposed to surrogate function modelling which imposes stronger limitations. Finding the optimum of a given function is thus equivalent to finding the multivariate distribution mean and covariance matrix. Parameter combinations are sampled from the distribution; each \textit{epoch}, a number of combinations (set a priori as the ``family size'', $N$) are generated, which are then evaluated - i.e. for each parameter combination, a corresponding simulation is executed and an error value computed. The error values are fed back to CMA-ES, which then improves the fit of the multivariate distribution; the covariance matrix is adapted - if relatively consistent responses are identified, the covariances are shrunk (i.e. the distribution ``zooms in'' on the minimum); the active component in the adaptation allows the covariances to enlarge in case better minima are identified further from the current estimate, or if an input is found to have no sensitivity on the output. Thus, the optimiser can adaptively ``zoom out'' from local minima, move towards global minima and ``zoom in'' on the best, physical parameters. This is also made possible by the addition of stochastic ``mutations'' to the generation of parameter combinations, which mimics the addition of new genetic material in a biological population, allowing it to escape local minima; this, naturally, would be impossible for gradient-based optimisers \cite{venter2010review}. In a comprehensive derivative-free optimisers benchmark, CMA-ES consistently ranked in the top ones for difficult black-box optimisation problems \cite{rios2013derivative}, hence our choice to include it in the ACCES ecosystem.

The only CMA-ES metaparameters that need to be set a priori are: i) the "family size", or the number of parameter combinations to evaluate in each epoch, and ii) the initial standard deviation, which sets the spread of the parameter combinations in the initial generation. The ACCES library adapts CMA-ES for large-scale, parallel evaluation of complex, time-intensive simulations (i.e., the function evaluations). To allow for maximum parallel execution efficiency, the family size is {often} simply set to the number of processors on the system, ranging from 8 in contemporary consumer workstations with multi-threaded CPUs to hundreds in expansive supercomputing clusters, such that each simulation trial has dedicated resources to run in parallel to the others; {note that the user sets this parameter based on the computational resources available or CPU usage desired}. 
A larger family size permits a more thorough exploration of the parameter space within a single epoch, achieving the optimum solution in fewer epochs, albeit at the cost of running more simulations in parallel. In our studies, the choice of family size impacted only the runtime of the optimisation process; the physical, globally-optimal results have consistently been correctly identified regardless of the family size.

To summarise, the CMA-ES optimiser provides a number of advantages making it uniquely suitable for our calibration problems \cite{rios2013derivative}:

\begin{itemize}
    \item Consistently finds global optima in difficult, noisy problems.
    \item Requires very few function evaluations ($\mathcal{O}(10-100)$ per free parameter) relative to other global optimisers, especially to other evolutionary algorithms.
    \item Only has two metaparameters, which are straightforwardly set to default values for the given problem by ACCES, while not affecting the success of the calibration, only its runtime.
    \item Does not require an initial guess, and is not sensitive to the initial metaparameters in our bounded calibration problems.
\end{itemize}

Its main disadvantages, or missing features, are solved by the ACCES implementation:

\begin{itemize}
    \item Scalarising multiple objectives. ACCES offers a scalariser interface for implementing specific methodologies such as multiplication and weighted addition.
    \item Inputs of different orders of magnitude are not natively handled.
    \item Cannot natively reload previous simulation runs, or historical data.
    \item Cannot save its optimisation state.
\end{itemize}

The last three are detailed in the following subsection.

\subsubsection{Parameter Scaling and Saving Simulation / Optimisation History}

As CMA-ES assumes a single value for the diagonal entries of the covariance matrix, it requires input parameters be of similar orders of magnitude. To circumvent this, ACCES scales each parameter down by $40\%$ of its accepted range (parameter bounds are given a priori), such that samples drawn from a normal distribution with unity standard deviation would completely cover the entire range. {For example, if the ranges for two parameters are $A \in (500, 1500)$ and $B \in (0, 1)$, then scaling each down by $40\%$ of their range (i.e. scaling factors $S_A = 400$, $S_B = 0.4$) would result in the scaled parameters $\overline{A} \in (1.25, 3.75)$ and $\overline{B} \in (0, 2.5)$ of equal ranges.}
Thus, CMA-ES can work with a phenotype space of unity starting standard deviations in the covariance matrix, generating parameter combinations which are then transparently up-scaled by ACCES and passed to the simulations during the parallel evaluation stage of each epoch.

ACCES replaces the internal CMA-ES random number generator with its own, seeded generator, such that sample generation becomes deterministic. As each epoch is evaluated, ACCES saves its entire simulation and optimisation state in a directory named ``access\_seed\{Seed Number\}'' (e.g. ``access\_seed123'' if the random seed were 123){; determinism for the calibration run is powerful for its ability to ``re-tread'' the optimisation steps and resume calibration at any given point should it crash (more details in the next paragraphs), or restart and let it run for longer from the final step. The simulation directory can be archived, transferred and inspected during and after calibration; it includes the following:}

\begin{itemize}
    \item The user-script modified at runtime via metaprogramming.
    \item The ACCES setup, as a TOML-formatted dictionary, containing relative paths within this directory, parameters, population size, target uncertainty, random seed, etc.
    \item Scaled and unscaled estimates, one for each completed epoch, as CSV files.
    \item Scaled and unscaled parameter combinations and their corresponding evaluated outputs, as CSV files.
    \item A log, as a text file, of the ACCES runs (re)started.
    \item A README, formatted as a Markdown file, of the directory contents.
    \item A directory of logged outputs, as individual text files piped from each simulation executed in a separate OS process.
    \item A directory of simulation outputs -- namely the ``error'' and ``extra'' variables defined in the user-script -- saved Python PICKLE binary files, which can encode arbitrary Python objects; one for each simulation. 
\end{itemize}

If an OS process running a simulation finishes, but no corresponding ``error'' file is found by the ACCES main driver code, it means that the simulation crashed early; ACCES notifies the user as the simulation is running, and all error messages are logged in the outputs directory for inspection. Depending on the simulation, it is possible that some parameter combinations are simply unfeasible, and some simulations are expected to crash, and their corresponding parameters should simply be avoided in the future; ACCES graciously handles such crashing simulations transparently. Crashing simulations' error values are simply set to Not-a-Number (NaN) floating-point representations, and CMA-ES avoids them in future epochs. However, it is also possible that an error in the user script results in all simulations crashing; if two consecutive epochs are filled with NaNs, ACCES finishes early and notifies the user of perpetually-unstable simulations. The ACCES directory contents are updated at the end of every epoch; it can be archived and copied to different machines for inspection even while calibrations are still running. 

As simulations are computationally expensive, if the main ACCES driver code crashes -- e.g. if the time allocated on a cluster job ends, or the workstation restarts -- it is important that calibration can be restarted and continue from its last point. CMA-ES follows a very strict sampling procedure to maintain its high convergence rates, and therefore the only way to resume a previous optimisation run is to feed back the exact same results, in the same order. ACCES does this transparently if a previous ACCES directory is found at the start of its run. However, due to the down- and up-scaling of parameters, approximation errors are introduced, which, if fed back to CMA-ES, will result in different random number generator states, and diverge from previous runs. Therefore, raw, unscaled values given to and received by CMA-ES in previous runs must be saved to disk and used directly when injecting historical values back upon resuming in a new ACCES run. Again, ACCES handles this transparently to the user.

\section{Methods and Materials}

\subsection{The Discrete Element Method}

The Discrete Element Method (DEM) is a computational technique used to simulate the behavior and interactions of large numbers of particles within a system, allowing for the quantitative prediction of complex phenomena of granular materials or other assemblies of ``discrete bodies'' \cite{blais2019experimental}. It has been used successfully in an exceptionally wide variety of fields, including i) geotechnical engineering, to study the behavior of soil, rock and snow in a variety of loading conditions, such as earthquakes and avalanches \cite{donze2009advances, bi2019effects, gaume2015modeling}, ii) the pharmaceutical industry for investigating and optimising API and excipient mixing and compaction, film coating, and tablet swelling and dissolution \cite{ketterhagen2009process, yeom2019application, kimber2012modelling}, iii) agricultural engineering, for modelling seeding, harvesting, soil-machinery interaction, and processing of agricultural products \cite{zhao2021applications, yan2022review}, iv) mining and mineral processing, for rock cutting, crushing, conveying and mineral composition predictions \cite{rojek2011discrete, li2014discrete, he2023correlations, shi2022recent}, v) astrophysics, for granular dampers in microgravity, rubble pile morphology on comets, Mars rover wheel performance, investigating the behavior of regolith and boulder excavation on asteroid surfaces \cite{sack2020granular, tancredi2012granular, zhang2020tidal, johnson2015discrete, kulchitsky2014discrete, brisset2022asteroid}.

In the 1970s P. Cundall and O. Strack developed a computer model for the simulation of rock assemblies by solving Newton's equations of motion for multiple spheres through the inclusion of \textit{contact forces}, acting at the intersection between two bodies and proportional to the overlap depth between them, marking the birth of the \textit{Discrete Element Method} \cite{cundall1979discrete}. Fundamentally, it is inspired by the older Molecular Dynamics (MD) technique \cite{hansson2002molecular, thompson2022lammps}, but including and focusing on \textit{stateful contact} between a large number of complex-shaped objects with rotational degrees of freedom; {contacts are considered stateful as they need to keep track of frictional history for the purposes of modelling hysteresis, which in turn requires significant software implementation differences.} 
It differs from the more recent rigid body dynamics technique (RBD) that is popular in robotics, films and video games by allowing inter-object penetration, which is used for complex but accurate contact behaviour; on the other hand, RBD focuses on simulation speed over microscopic accuracy, and the inclusion of kinematic joints and movement constraints \cite{borisov2018rigid}.

Simulating a granular assembly using DEM is, fundamentally, solving a (large) system of differential equations; for each particle, the following holds via Newton's translational and Euler's rotational equations of motion:

\begin{equation}
\centering
\begin{split}
     \frac{d}{dt} \begin{bmatrix} X(t) \\ R(t) \\ P(t) \\ L(t) \end{bmatrix} = \begin{bmatrix} V(t) \\ \omega^* (t)* R(t) \\ F(t) \\ \tau (t) \end{bmatrix}
\end{split}
\label{eq:dem_eqs}
\end{equation}

Where $X$ is the 3D position, $R$ is a rotation matrix defining the orientation of an object relative to the system coordinates, $P$ is the particle momentum (i.e. mass multiplied by velocity), $L$ is the angular momentum (i.e. moment of inertia multiplied by angular speed), $V$ is the particle velocity, $\omega^*$ is the skew-symmetric matrix constructed from the angular velocity, $F$ is the sum of forces acting on the particle and $\tau$ is the sum of torques acting on the particle. Notice how a discrete element state at any given time step is fully defined by the differentiated vector. 

A typical DEM simulation involves the following general steps:

\begin{enumerate}
    \item Simulation set up: \begin{enumerate}
        \item Model the 3D geometry of the system considered and load it into the simulation domain.
        \item Select the object materials or \textit{species} that will be considered.
        \item Define a number of forces acting on each material (e.g. external forces such as gravity) and pairs of materials (i.e. \textit{contact forces}).
        \item Insert a given number of particles of some size and material.
        \item Optionally, set particle insertion regions and particle deletion boundaries.
    \end{enumerate}
    
    \item Running the simulation: \begin{enumerate}
        \item Compute all the forces and torques acting on each particle by summing each individual contribution (e.g. external forces, contact forces and torques, etc.) and calculate the right-hand side of Equation (\ref{eq:dem_eqs}).
        \item Move the simulation forward in time - that is, predict the next vector value from the current derivative - using some integration scheme (e.g. velocity Verlet, Gear predictor-corrector, etc.).
        \item Optionally, use and maintain a contact resolving acceleration data structure, such as a neighbour list or a bounding-volume hierarchy.
    \end{enumerate}
    
    \item Visualisation and post-processing: \begin{enumerate}
        \item The saved particle locations, orientations, velocities, forces and torques at every time step can be visualised using specialised software (e.g. ParaView \cite{fabian2011paraview}).
        \item Other system-specific data can be extracted using post-processing routines (e.g. velocity vector field, dispersion, etc.).
    \end{enumerate}
\end{enumerate}

\subsubsection{Software Used}

In this work, the LIGGGHTS open-source DEM engine is used for its best-in-class simulation speed, scalability (based on the LAMMPS molecular dynamics architecture, which it extends), wide validation and industrial adoption \cite{kloss2011liggghts, thompson2022lammps, kloss2012models}. Open-source digital models of the experimental systems described in Section \ref{sec:experimental_systems}, developed by the authors, were used. Key DEM parameters used throughout the paper are given in Table \ref{tab:dem}.

\begin{table}[htbp]
\centering
\caption{Key DEM parameters used for the granular simulations in this paper.}
\label{tab:dem}
\begin{tabular}{ll}
\textbf{Timestep}         & 1e-5 s                        \\
\textbf{Shape Model}      & Spherical                     \\
\textbf{Contact Model}    & Hertz-Mindlin with            \\
\textbf{}                 & CDT Rolling Friction          \\
\textbf{Young's Modulus}  & 1e6 Pa                        \\
\textbf{Poisson Ratio}    & 0.4                           \\
\textbf{Sliding Friction} & Calibrated (see next section) \\
\textbf{Rolling Friction} & Calibrated (see next section) \\
\textbf{Restitution}      & Calibrated (see next section) \\
\textbf{Saving Format}    & VTK                          
\end{tabular}
\end{table}

{
Note that the same coefficients have been used for both particle-particle and particle-wall interactions.

The value of the timestep chosen corresponds to $\sim 11\%$ of the Rayleigh timescale, as computed for the smallest particle diameter, which is small enough to guarantee stable integration accuracy \cite{otsubo2017empirical}. The Young's Modulus and Poisson ratio values correspond to softened particles, which make the ODE problem (Equation \ref{eq:dem_eqs}) significantly less stiff; a commonly-employed technique in DEM, as long as they are large enough, they do not affect the modelling of bulk DEM dynamics - the specific values used are typical in the DEM community \cite{coetzee2017calibration}.
\blue{The fastest-moving system in the present study is the ResoDyn acoustic mixer (see next section) which vibrates at an amplitude of $6.69$ mm at a frequency of $62.44$ Hz, resulting in the largest velocity in the case of simple harmonic motion of $2.62$ m/s; a timestep of $10^{-5}$ s results in $0.0262$ mm maximum displacement per timestep, which is over 38 orders of magnitude smaller than the smallest particle diameter, and thus is small enough to accurately resolve the particles' motion.}

}

\subsection{Experimental Systems}
\label{sec:experimental_systems}

The materials, experimental systems, and imaging instruments employed in this work are detailed in this section. {Among the instruments modelled using DEM, note that while the rotating drum is used for both calibration -- deliberately using simple, readily-available optical measurements from the instrument itself for wide accessibility -- and simple validation, the tumbling mixer and vibrofluidised bed are used solely for the validation of the calibrated DEM particle properties in systems of different scale and operating regime (full details below), and employ extensive testing using positron emission particle tracking (PEPT).}

\subsubsection{Powder Modelled}

\begin{figure*}[htbp]
    \centering
    \includegraphics[width=\linewidth]{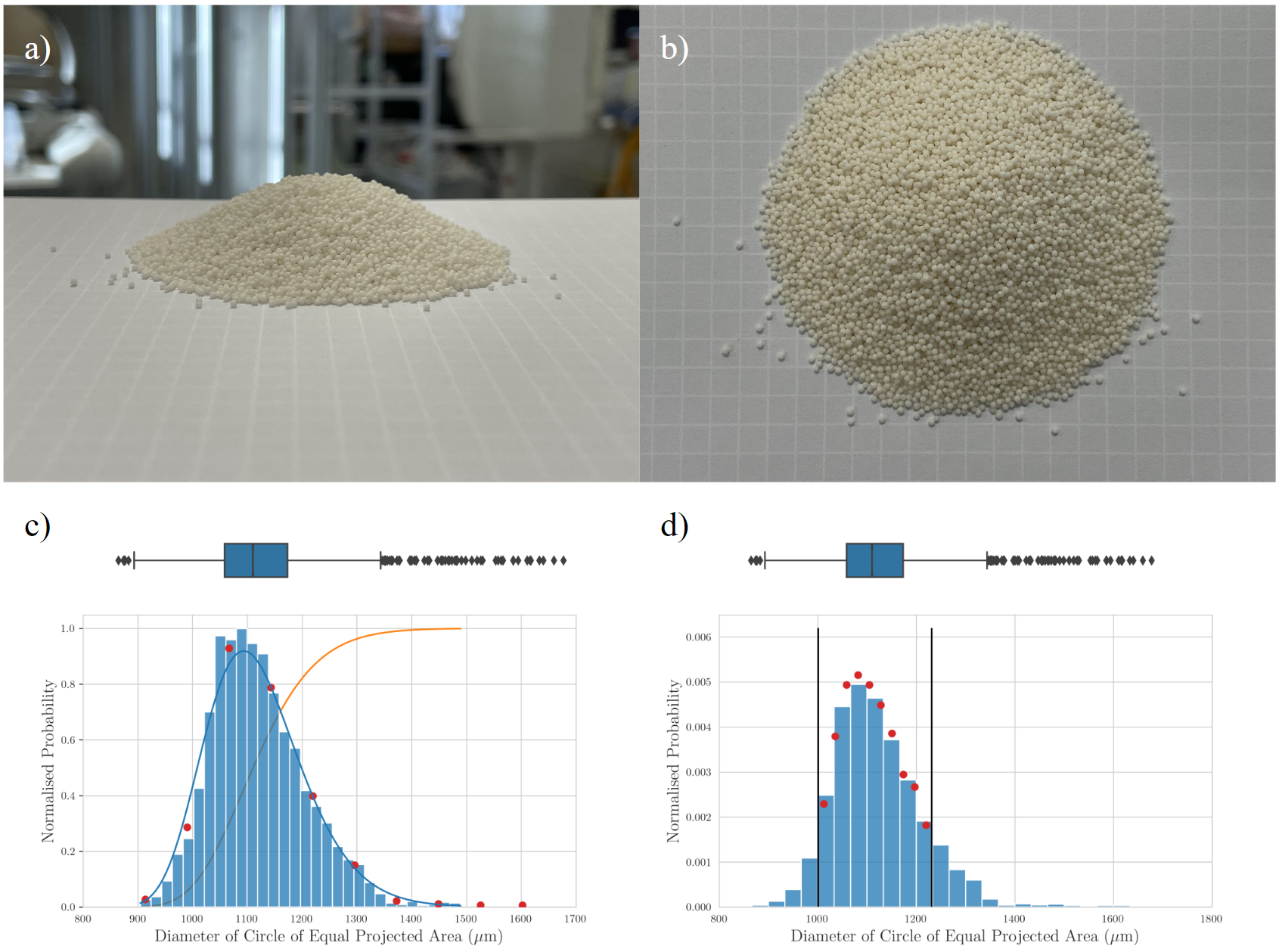}
    \caption{MCC particles utilised for the DEM application of ACCES; panels $a)$ and $b)$ display side and top-down photographs against 5 mm square lattice paper, respectively. Panel $c)$ depicts the particle size distribution (PSD), as imaged using a Sympatec QicPic. This includes the raw histogram (blue vertical bars), the cumulative distribution (orange line in the background), the fitted log-normal distribution (blue line), and an example of 10 size fractions derived between the 0.001 and 0.999 quantiles (red circles). The upper box plot details the first quartile (left box edge), the median (central vertical line), the second quartile (right box edge), the Tukey fences (left and right whiskers calculated as $Q_1 - 1.5 IQR$ and $Q_3 + 1.5 IQR$ where $IQR = Q_3 - Q_1$ represents the inter-quartile range), and outliers (black diamonds). Panel $d)$ shows black vertical lines that mark the boundaries of the ``log-cut'' PSD employed in the DEM simulations, where the natural logarithm of the PSD is taken and lower/upper limits are computed as $\pm 1.281552 \sigma$ around the median (still in log space), before being transformed back into standard Cartesian space through exponentiation. The red circles indicate the 10 distinct size fractions selected within these limits for use in the simulations.}
    \label{fig:psd}
\end{figure*}

To focus on the extensive validation of the ACCES calibration architecture and the correct modelling of physics in DEM for an industry-relevant powder, microcrystalline cellulose (MCC) granules were used, which, while relatively free-flowing, still show a relatively large particle size distribution and variation in surface asperity. The absolute density used is 1580 $\mathrm{kg m^{-3}}$. Photographs of the particles relative to 5 mm square lattice paper is shown in panels $a)$ and $b)$ of Figure \ref{fig:psd}.

It is very common for DEM practitioners to up-scale, or ``coarse-grain'' particles, so that larger systems can be tractably simulated. These coarse-grained (CG) particles, therefore, effectively model \textit{parcels} of the original material granules; the degree of coarse-graining possible, while still retaining the dynamics of the system is strongly system-dependent, e.g. it was found that coarse-graining ratios of only 1.3 can manage to correctly model hoppers, while ratios of 9 and above can work for drums, as long as over 25 particles fit between the drum sides \cite{coetzee2019particle}. Coarse-graining methodologies remain an open, and active field of research in the DEM community.

One of the main challenges in up-scaling smaller powders to simulate larger systems is that the upper limit of the particle size distribution (PSD) often increases beyond what is physically suitable for the system. For instance, a powder with a PSD ranging from 100 to 500 microns may be too small to simulate in a rotating drum of 84 mm diameter and 20 mm thickness (discussed in the next Section). However, when this powder is coarse-grained by a factor of 6 (which is not a high number, see previous paragraph), the lower end of the PSD shifts to 600 microns, which remains relatively small compared to the system size, while the upper end now extends to 3 mm - this results in fewer than 7 particles fitting across the axis of the drum!

We have developed a ``log-cut'' method to set the particle size distribution: since the vast majority of PSDs in nature and industry adhere to log-normal distributions (as obvious in panel $c)$ of Figure \ref{fig:psd}), we select a segment of a PSD in log-space. After taking the natural logarithm of the PSD, we calculate the median $M_{log}$ and standard deviation $\sigma_{log}$ of the particle sizes. To encompass 80\% of a log-normal distribution, which appears normal in log space, we determine the bounds of our PSD in log space as $(M_{log} - 1.281552 \sigma_{log}, M_{log} + 1.281552 \sigma_{log})$. These bounds are then transformed back into normal Cartesian space to define the limits of our PSD as:

\begin{equation}
\centering
\begin{split}
    \textrm{min}_{PSD} = \exp \left[ M_{log} - 1.281552 \sigma_{log} \right]
    \\
    \textrm{max}_{PSD} = \exp \left[ M_{log} + 1.281552 \sigma_{log} \right]
\end{split}
\label{eq:logcut}
\end{equation}

Therefore, we accurately address the asymmetry typically observed in the PSD of powders that nearly follow a log-normal distribution. The derived log-cut PSD is depicted in the MCC distribution shown in panel $d)$ of Figure \ref{fig:psd}; the ten fractions shown as red circles were implemented in the DEM simulations. This approach also allows powders of various sizes to be coarse-grained by the same factor, preserving their volume ratios constant — and thus maintaining their physical segregation rates and degrees.

\blue{Note that for brevity, and in order to keep the present paper focused on i) the ACCES general-purpose calibration framework, ii) a DEM calibration methodology for dense granular flows, and iii) comprehensive validation of i) and ii) with PEPT in instruments spanning vastly different scales and operating regimes, we have deliberately focussed on a single, comparatively simple particulate medium. The framework has, however, also been tested for cohesive, angular, and elongated particles, as well as for binary mixtures of particles. Results obtained with these materials will be published in a series of future, related papers.}

\subsubsection{Granutools GranuDrum}

\begin{figure*}[htbp]
    \centering
    \includegraphics[width=\linewidth]{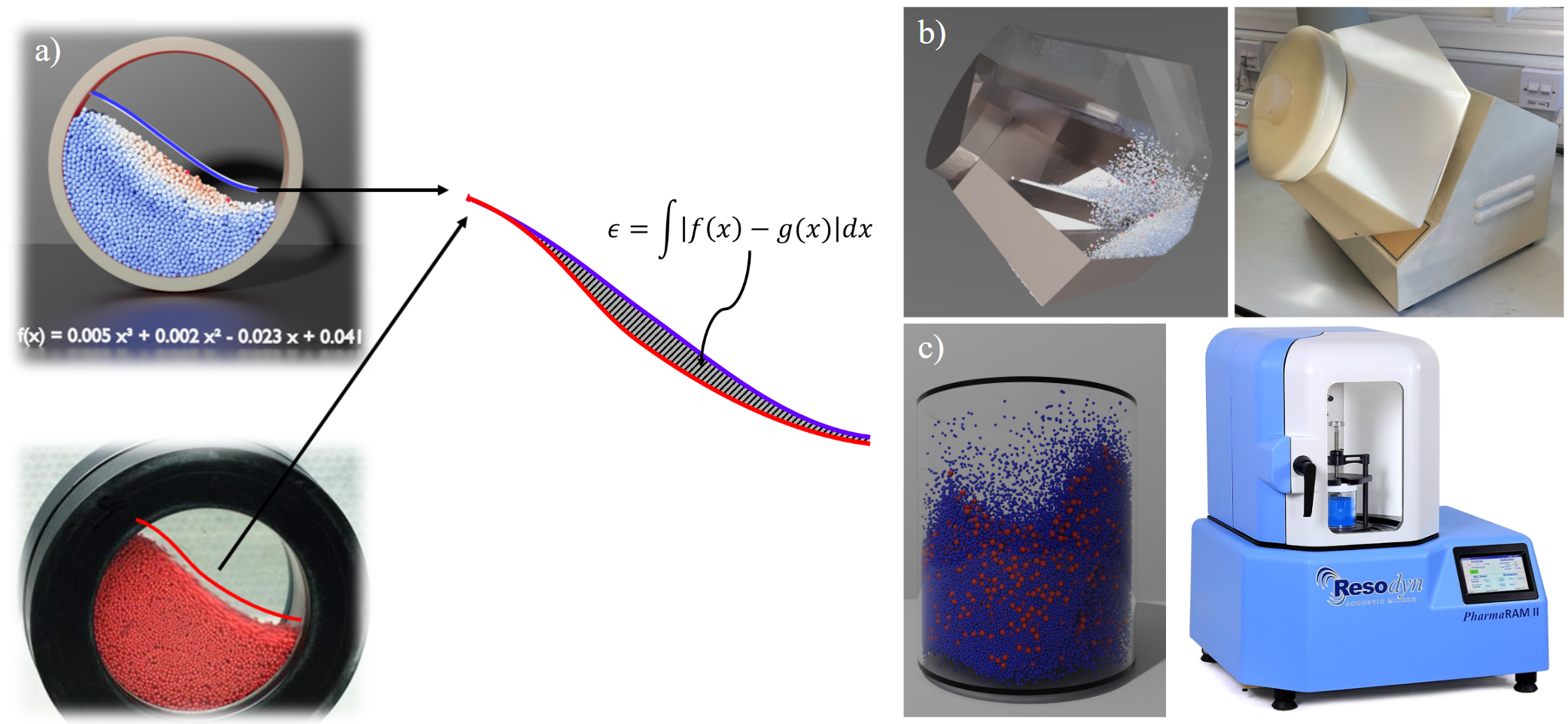}
    \caption{Discrete element method digital models used for i) particle contact parameters calibration using ACCES, with the Granutools GranuDrum experimental-simulational superimposed free surface error areas (i.e. the highlighted shaded area), at 15 and 45 RPM, acting as the error functions (panel $a)$), and ii) simulation validation using the Pascall tumbling mixer (panel $b)$) and Resodyn acoustic mixer (panel $c)$). Each panel shows both experimental photographs of the systems, as well as renders of their corresponding DEM digital models.}
    \label{fig:twins}
\end{figure*}

Shown in panel $a)$ of Figure \ref{fig:twins}, the Granutools GranuDrum is a commercially-available experimental powder characterisation instrument, offering a standardised granular rotating drum system, where a horizontal cylinder containing the sample material rotates at a controlled speed \cite{lumay2012measuring}. As the drum (diameter 84 mm, depth 20 mm) rotates, 50 mL of powder is subjected to a continuous flow under the influence of gravity, allowing the relative characterisation of the flowing behaviour of the material. The shape of the resulting free-flowing surface is closely-tied to the frictional and cohesive properties of the particles (and, to a lesser extent, their restitution), but with no constitutive law currently-available allowing us to directly measure these properties. This is a perennial problem in powder technology where, due to the complexity of bulk powder phenomena, the primary way to characterise granular materials is via bulk measurements like the dynamic angle of repose of the free surface shape of a rotating drum \cite{windows2016numerical}. This stands in stark contrast with other fields, where microscopic measurements are readily available and influence macroscopic behaviour predictably; a prime example thereof is the effect of viscosity in fluid dynamics.

Instead, ACCES allows us to \textit{back-compute} the particle properties which would result in a corresponding high-fidelity digital model to match experimental measurements (shown in panel $a)$ of Figure \ref{fig:twins}); thus, ACCES provides a way to numerically ``measure'' the microscopic particle contact parameters from a macroscopic image (full details in Section \ref{sec:methodology}).

\subsubsection{Pascall Mixer}

As depicted in panel $b)$ of Figure \ref{fig:twins}, the Pascall tumbling mixer involves an octahedral, baffled, double-sloped container rotating around an axis that is at 60 degrees from the horizontal. It is commonly used for lab-scale mixing of pharmaceutical excipients. Mixing 500 mL of powder, it operates at a scale that is an order of magnitude greater than the rotating drum (50 mL) and the Resodyn acoustic mixer (using 10 mL {here to also include a very small-scale benchmark}). Furthermore, due to the baffles lifting the powder before sliding and raining down into the bulk, it operates in a regime where both frictional ($\mu_s,\mu_r$) \textit{and} collisional ($\varepsilon$) parameters are expected to be important.

\subsubsection{Resodyn Acoustic Mixer}

Shown in panel $c)$ of Figure \ref{fig:twins}, the Resodyn acoustic mixer is an archetypal vibrofluidised bed that is popular in the pharmaceutical, battery and energetics industries for its high performance in powder, fluid and paste mixing, granulation, coating, and dissolution \cite{kline2022probing, yuksel2022resonant, rodriguez2023manufacturing, frey2024use, leung2014new}. 
The lab-scale RAM II used in the present study vertically shakes a 60 mm wide, 80 mm tall cylindrical vessel filled with 10 mL of powder at the resonant frequency of the system, \blue{where the potential energy stored in the springs can be most efficiently transferred onto the plate on which the cylindrical vessel is fixed - in the present case around 62.44 Hz - thus drastically reducing energy requirements compared to excitation under non-resonant conditions. Compared to the mass of material in the vessel, the vibrating plate is significantly heavier, thus effectively operating in a \textit{constant energy input} regime with respect to the powder}. In systems such as the RAM, the vibration amplitude has a strong and direct effect on the dynamics of the system, and thus its mixing strength \cite{rosato2020segregation}; in this case, the highest setting used resulted in 6.69 mm amplitude, {equivalent to a dimensionless acceleration of 100 G}. Again, the system operates at an order-of-magnitude difference in scale from the other validation systems. Importantly, the Resodyn acoustic mixer -- unlike the rotating drum used for calibration -- operates in a distinctly collisional regime, where the restitution is expected to be the dominating particle property, thuis representing a deliberately challenging validation system. More generally, the restitution coefficient is infamously difficult to calibrate from dense granular systems, such as shear cells, angle of repose testers or rotating drums \cite{windows2022calibration}. As shown in Section \ref{sec:methodology}, this is indeed possible if the entire drum free surface shape, along with its edges, is included - still, the restitution is the last parameter to be calibrated, and mainly becomes apparent once the bed expansion (controlled by the number of particles, sliding and rolling friction) is calibrated.

\subsubsection{Positron Emission Particle Tracking}
\label{sec:pept}

\begin{figure*}[htbp]
    \centering
    \includegraphics[width=0.9\linewidth]{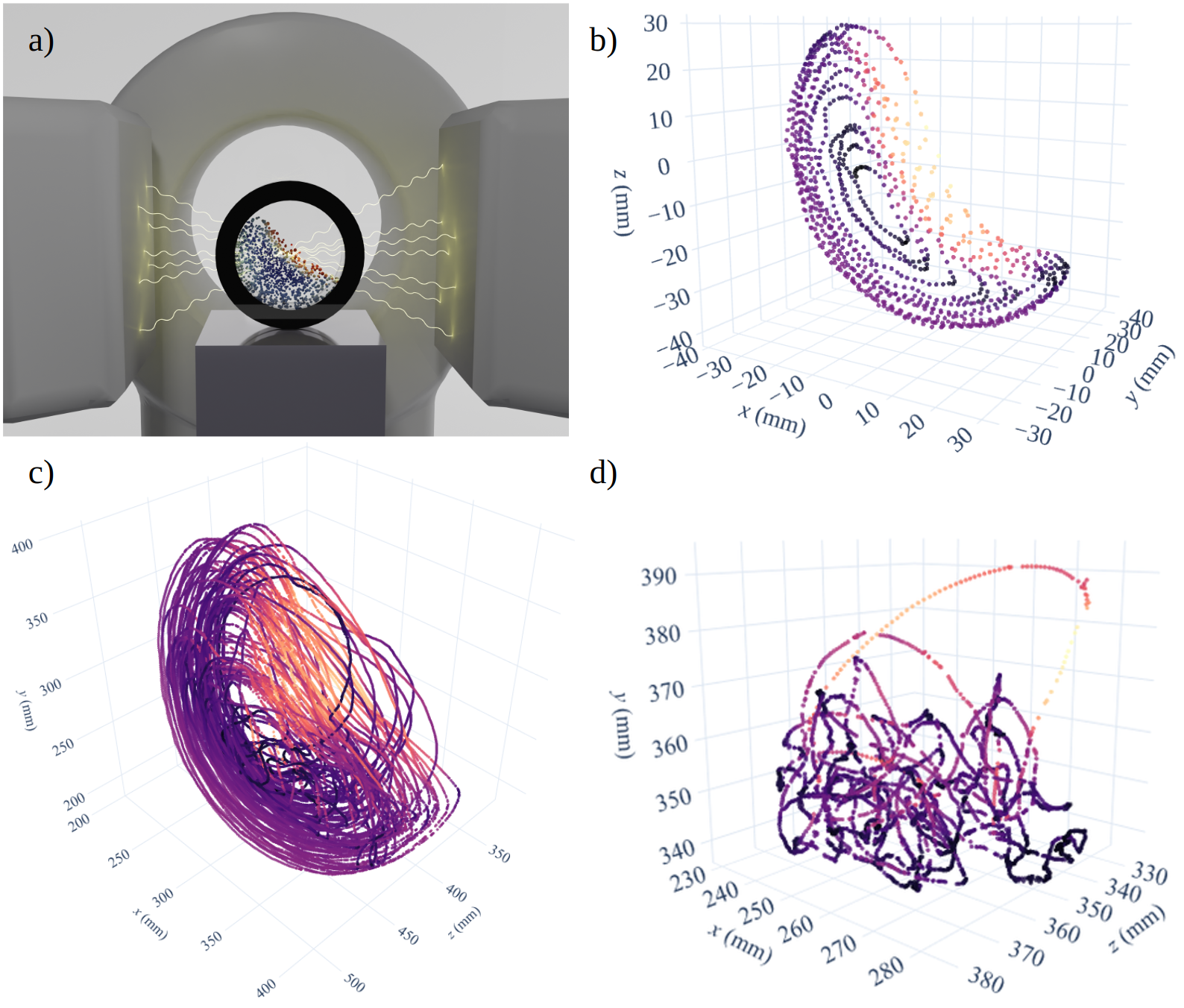}
    \caption{Positron emission particle tracking (PEPT) as used for experimentally investigating the calibration and validation systems in this paper, yielding particle trajectory data similar to the discrete element method simulations. Panel $a)$ illustrates the emitted gamma rays by a radioactively-labelled tracer particle within a system of interest, which are captured by gamma camera detectors. Panels $b)$, $c)$ and $d)$ show short tracer trajectories, as outputted after processing the gamma rays using the PEPT-ML algorithm, for the rotating, tumbling mixer, and the vibrofluidised bed, respectively.}
    \label{fig:pept}
\end{figure*}

Positron emission particle tracking (PEPT) is a technique that allows the three-dimensional imaging of large, optically-opaque industrial systems with sub-millimetre and sub-millisecond spatio-temporal resolution \cite{windows2020positron,windows2021recent}. Due to the use of highly-penetrating 511 keV gamma rays emitted, PEPT can be used to probe the internal dynamics of dense industrial equipment. Radioactively-labelled tracers can be produced from a variety of materials such as glass, graphite, silica or resin, from micron- to centimetre-scales in liquid, gaseous or particulate media \cite{windows2022positron}. PEPT has been used to investigate industrial systems including vibrated and fluidised bed reactors \cite{martin2005hydrodynamic, viswanathan2006comparison, seiler2008statistical}, rotating drum systems \cite{gonzalez2015forced, windows2016understanding} or a variety of mixers \cite{marigo2013application, guida2010pept}, with current research pushing for the use miniaturised tracers in biological applications \cite{langford2017three, chang2020novel}. Lagrangian particle tracking in general can offer a wealth of information about the inner dynamics of a system, from residence time and velocity distributions, to dispersion and granular temperatures; the ability to compute these properties with high resolution \textit{in situ} and \textit{in vivo} has increased the popularity of PEPT in recent years, as it left its birth place in Birmingham to be extensively used and improved in new centres around the world, including South Africa and the US \cite{windows2022positron}.

In the present work, a non-invasive tracer -- simply an MCC particle taken from the bulk powder -- is created by labelling its surface with Fluorine-18 beta-emitting radioisotopes, produced by the cyclotron of the University of Birmingham Positron Imaging Centre. This tracer undergoes $\mathcal{O}(10^7)$ radioactive disintegrations per second, each producing a positron, which near-immediately annihilates with a nearby electron, thus producing a pair of anti-parallel gamma rays, as illustrated in panel $a)$ of Figure \ref{fig:pept}. These gamma rays are captured by a gamma camera, in this case the ADAC Forte parallel-screen detector at Birmingham \cite{herald2021monte}, and saved by a digitizer as binary files. The gamma rays are then processed in overlapping samples of \~200 at a time, each sample being used to triangulate the tracer location; processing the whole dataset yields continuous tracer trajectories, sampled hundreds of times per second. Using the PEPT-ML algorithm \cite{nicucsan2020positron}, the accuracy of the resulting trajectories is improved beyond the native hardware capabilities of the detectors. Short trajectory segments for the calibration and validation systems considered in this paper are depicted in panels $b), c), d)$ of Figure \ref{fig:pept}.

Finally, the hours-long tracer trajectories are used to compute time- and/or length-averaged statistics about the system being studied, including velocity distributions \cite{windows2013boltzmann}, granular temperature fields \cite{windows2013thermal}, diffusion coefficients \cite{parker1994industrial,lee2012twin,windows2014self}, segregation patterns \cite{windows2014effects} and three-dimensional density distributions \cite{hensler2015positron}. These allow direct, 1-to-1 comparison with equivalent statistics extracted from DEM simulations for the comprehensive validation of the internal dynamics of the systems modelled; they are detailed in Section \ref{sec:validation}.

\section{Results and Discussion}

{This section describes the application of ACCES to the calibration of particle properties in discrete element method digital models (Section \ref{sec:methodology}), followed by extensive validation with quantities not directly included in the calibration, in systems spanning orders of magnitude differences in scale (Section \ref{sec:validation}).}

\subsection{Discrete Element Method Calibration}
\label{sec:methodology}

    In order to correctly calibrate the microscopic parameters of a physical simulation against macroscopic experimental measurements, the calibration targets must be i) sufficiently sensitive to the inputs, and ii) provide enough ``closure'' to separate out their effects. While some degree of confounding between inputs is expected -- and in many cases impossible to avoid -- {the collection of calibration measurements used must be comprehensive enough that the input parameters can be ``fixed'' to unique values; in other words, there is a single combination of input values that reproduce all calibration measurements used}. 

    As explained in Section \ref{sec:ACCES}, multiple calibration targets may be correctly scalarised into a single objective function. {In mathematical optimisation, a Pareto front is a set of input values combinations in a multi-objective optimisation scenario that represents the best possible trade-offs between different competing objectives, where no point on the front can be improved on one objective without worsening at least one of the other objectives \cite{van1998evolutionary}. Following from the discussion above on the necessary uniqueness of DEM particle properties that reproduce the calibration measurements}, as physically-correct calibration must not yield a Pareto front, multiple objectives may safely be collapsed onto a single one. 
    Discrete element method simulations of granular systems are particularly nuanced in terms of calibration. For example, in some situations the effects of particle sliding friction (opposition to translation) may be `traded off' for rolling friction (opposition to rotation); that is to say, the same bulk behaviour may be produced by two (or more) non-identical combinations of sliding and rolling friction. It is likely in such cases to identify some set of parameters that reproduce the \textit{calibration} measurement precisely, but fail to quantitatively faithfully reproduce the dynamics of the same particles in another (\textit{validation}) system \cite{windows2022calibration}.

    In other words, similar to how $N$ variables require $N$ mathematical closures to be solved for, calibration may be summarised as ``$N$-parameters, $N$-tests''; however, this does not necessarily require $N$ distinct characterisation instruments. The rotating drum free-flowing surface (shown in panel $a)$ of Figure \ref{fig:twins}) used in this work, for example, can be approximated by a 3\textsuperscript{rd} order polynomial; disregarding the polynomial coefficient that represents the bed height (i.e., the zeroth order coefficient), this equates to having three closure relations corresponding to the linear, squared, and cubic terms of the polynomial. Nonetheless, it is still possible for one particle parameter to be traded-off for another, particularly with sliding and rolling friction. Relying solely on a single rotating drum test for calibration can lead to non-unique solutions or ``overfitting'' of the free surface.

\begin{figure*}[htbp]
    \centering
    \includegraphics[width=0.9\linewidth]{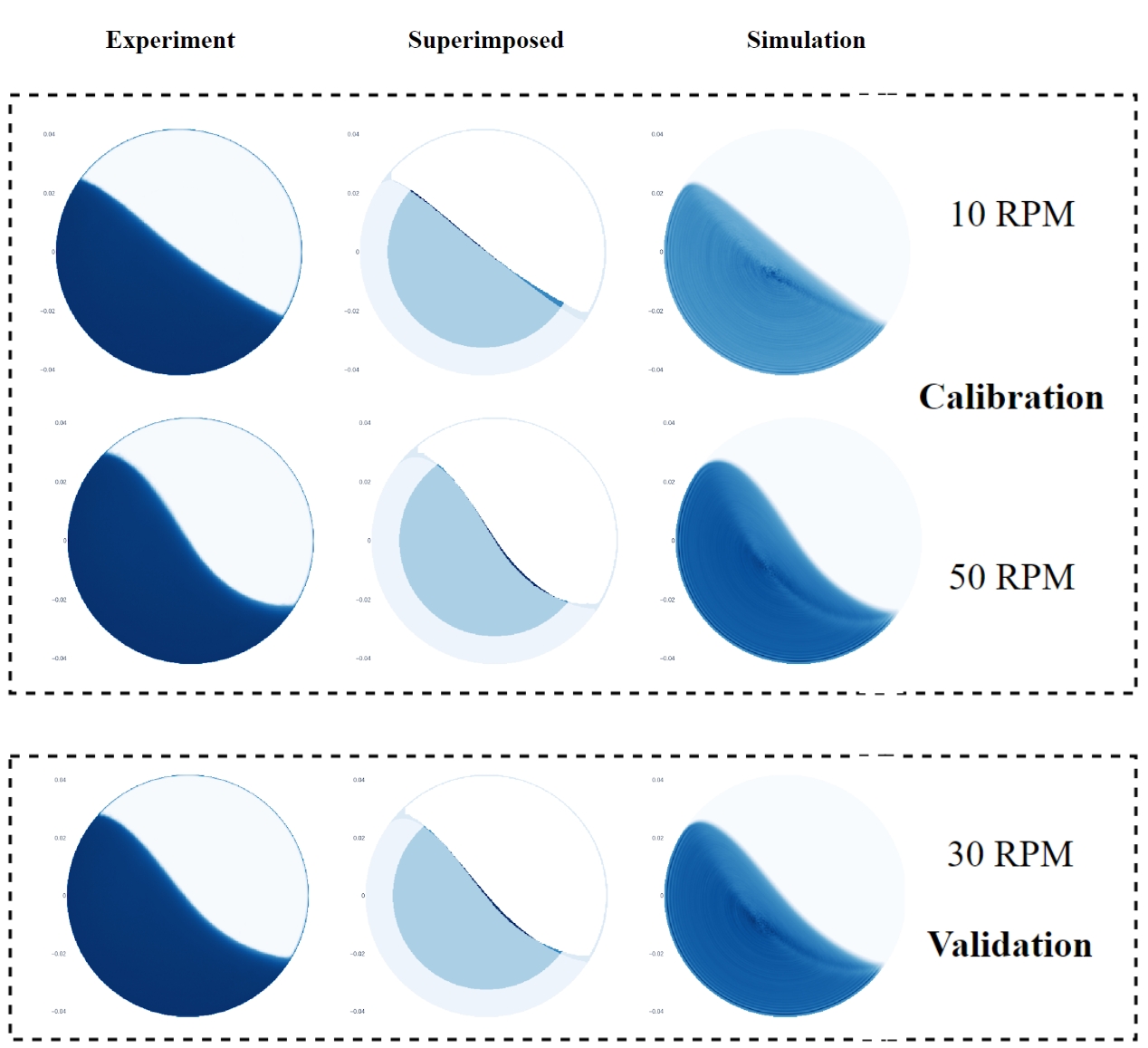}
    \caption{Illustration of ACCES-based calibration of DEM simulations: for each ACCES-predicted trial parameter combination, two DEM simulations of a rotating drum are conducted, at 10 RPM (top row) and 50 RPM (middle row). For each rotation speed, the simulated bed surface (depicted in the right column) is compared with the experimental free-flowing surface image (shown in the left column). Superimposing the two images (middle column), we calculate the total area error where the surfaces do not align (indicated by a dark blue thin layer). The errors calculated for 10 and 50 RPM are then multiplied together to produce a single scalar error value to be minimised via ACCES optimisation of the particle contact parameters. \blue{The simulations are coloured by the density of a 2D superposition of particles in the vertical plane. A validation run can be conducted at a different RPM to the ones used for calibration (the bottom row shows such a validation run at 30 RPM) ensuring that the system is accurately, physically characterised and not ``overfitting'' the specific measurements included in calibration.}}
    \label{fig:superimposed}
\end{figure*}

    A particularly powerful DEM calibration methodology developed in this work involves the use of two different rotation rates for the rotating drum, specifically when they correspond to distinct operating regimes. As shown in Figure \ref{fig:superimposed}, free-flowing surface shapes for both 10 RPM and 50 RPM cases are used for calibration, the two operating distinctly in the rolling and cascading regimes, respectively. It is ``difficult'', in a calibration sense, to reproduce both systems simultaneously, thus resulting in a unique, physical solution for the calibration, as detailed and extensively validated in the following sections.

    \blue{While measurements of the rotating drum free surface at two distinct RPM provides a suitable number of comparators to calibrate the 4 free calibration parameters discussed in the present paper, if one were to calibrate a significantly larger number of parameters, the problem would no longer be well-constrained, and thus the user could not be certain that the results offered represented a unique solution (i.e.\ the `true' values of the calibrated parameters). This problem can very easily be overcome, however, through the introduction of additional measurements using other suitable tools (or indeed additional measurements, such as the cohesive index, using the same tool), until a suitable number of constraints can be imposed for the particular calibration problem at hand. Naturally, increasing the number of free parameters to be calibrated will increase the number of function evaluations required, and thus the duration of the calibration run. However, prior studies using ACCES \cite{herald2022autonomous,hart2024autonomous} empirically suggest a sub-linear scaling of the required number of function evaluations with the number of additional parameters, meaning that the calibration problem is likely to remain tractable even when calibrating a large number of parameters. Nonetheless, as discussed in greater detail in prior works \cite{windows2022calibration}, it is typically advisable to perform a screening step before running ACCES -- i.e.\ to directly measure those parameters (such as PSD) which are easily measurable, and only calibrate those parameters which need calibrating}  

    \blue{
    It is also worth noting that while calibration using the rotating drum will successfully reproduce the internal dynamics of the vibrofluidised bed in Section \ref{sec:validation_resodyn}, this is not a commutative property - that is to say, a validation system cannot necessarily serve as a calibration instrument. In this example, vibrofluidised beds are particularly insensitive (or at least, only indirectly sensitive) to the frictional properties of powders when operating within the gaseous regime \cite{windows2013thermal,windows2014effects,windows2016convection} that would otherwise be essential for the internal dynamics of rotating granular drums. To put it simply: data obtained from a rotating drum are sufficient to calibrate a vibrofluidised bed, but those obtained from a vibrofluidised bed are not sufficient to calibrate a rotating drum (or indeed other dense, frictional systems).

    }

    To summarise, the Discrete Element Method calibration methodology developed using ACCES for dynamic granular systems involves the following:

    \begin{enumerate}
        \item \textbf{Input parameters}: sliding friction $\mu_s \in (0, 1)$, rolling friction $\mu_r \in (0, 1)$, restitution $\varepsilon \in (0.05, 0.95)$ and number of particles $N_P$, determined as 75\%-125\% of the number of particles found in 50mL in a cylindrical beaker simulation.
        \item \textbf{Optimisation target}: free-flowing surface error areas at 10 RPM and 50 RPM, multiplied together.
        \item \textbf{ACCES parameters}: the only meta-parameter to set is the \textit{family size}, i.e. the number of simulations to run in parallel per epoch; as detailed in Section \ref{sec:ACCES}, this only affects the optimisation time, and not the result. To simulate a typical consumer-available machine, it was set to 8, corresponding to the number of cores on an ordinary CPU.
    \end{enumerate}

    Importantly, in addition to calibrating common contact parameters such as sliding friction, rolling friction, and restitution, the number of particles $N_P$ included in the drum is also an optimisation input. The powder frictional properties influence its bulk density, and hence the particle bed expansion inside the drum, which differs based on its rotation rate and operating regime. As a consequence, we can separate the effects of friction from those of bed expansion -- and thus, alongside the two free surface shapes, we are indirectly calibrating against, and reproducing the powder bulk density; finally, this further enhances the sensitivity of the two-RPM rotating drum-based calibration method.

\subsubsection{Implementation using ACCES}

    The specific ``ACCES simulation script'' -- the user code in panel $b)$ of Figure \ref{fig:architecture} -- computing a calibration error value for one set of input parameters, is a simple Python script doing the following:

    \begin{enumerate}
        \item Creates a copy of the template DEM simulation directory -- which contains the drum STL meshes and two LIGGGHTS input scripts, each defining and running a simulation, at 10 RPM and 50 RPM, respectively -- and modifies the LIGGGHTS input scripts to use the given parameter combination (as defined by the ``parameters'' variable).
        \item Launches the two rotating drum simulations for the given parameters as separate OS processes -- each running the ``liggghts'' executable -- concurrently, then waits for them to finish.
        \item Post-processes the generated simulation outputs, from the raw VTK files of the particle positions at 100 frames per second, into rasterized 2D images, as done by a camera photographing the powder flow, using the KonigCell library \cite{Nicusan_KonigCell_Quantitative_Fast_2022}.
        \item For the 10 and 50 RPM simulations, the number of pixels not matching the corresponding binarised experimental images is multiplied by the physical area spanned by a pixel, to yield the error areas. The ``error'' variable is the simply set to a list containing the two individual error areas.
    \end{enumerate}

    Such workflows are already common in numerical modelling studies, including being implemented by DEM practitioners, albeit often organised in separate files or directories, and executed manually. ACCES-based calibration simply requires that they can be launched from a ``main orchestrator script''.

\subsubsection{Converging on the Physical Parameters}
\label{sec:converging}

\begin{figure*}[htbp]
    \centering
    \includegraphics[width=\linewidth]{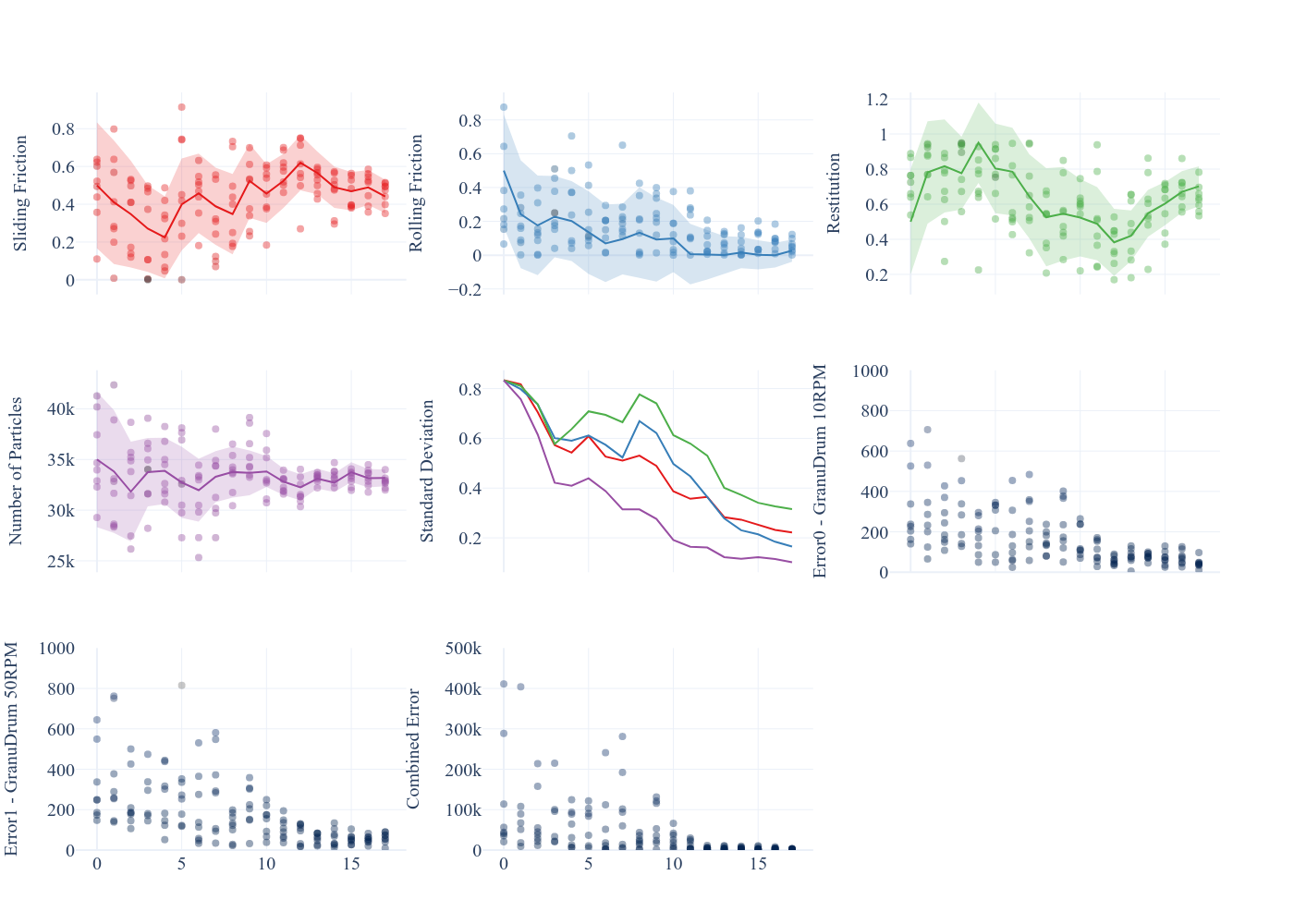}
    \caption{``ACCES convergence plot'', showing the contact parameters being calibrated by ACCES and minimising the flowing-surface area errors for the two rotating drums. Each vertical line on the X-axes represents an optimisation epoch, during which 8 parameter combinations are trialled concurrently. Each circle denotes a parameter combination tested by ACCES. For each combination, two rotating drum simulations are conducted, at 10 and 50 RPM. The standard deviation panel shows the uncertainty associated with each parameter (color-coded to match its respective subgraph) and the search area. Calibrating these parameters is equivalent to reducing this uncertainty. The solid lines in each parameter subgraph indicate the optimal value estimated at each epoch, while the height of the confidence band corresponds to the uncertainty of the parameter.}
    \label{fig:convergence}
\end{figure*}

For each batch of 8 parameter combinations produced by ACCES -- referred to as an ``epoch'' - the simulation script is executed in parallel, with each parameter combination. The execution is done following the scheduling interface -- e.g. on separate OS processes, or independent SLURM jobs (full details in Section \ref{sec:ACCES}). After waiting for all jobs to finish, ACCES collects the error values saved by these simulations and generates another set of 8 parameter combinations for the next epoch. This process repeats iteratively, improving the accuracy of the parameter estimates and converging towards an optimised set of input values that minimise errors for the drum at both 10 and 50 RPM. As indicated in the uncertainty graphs (the fifth subplot in Figure \ref{fig:convergence}), the first parameter typically calibrated is the number of particles, indicated by the rapid descent of the purple solid line, due to its significant impact on the bed volume, and hence error area. This is followed by the calibration of sliding and rolling frictions, in red and blue. The last parameter to be calibrated is the restitution coefficient, in green, as its influence is subtler and only becomes noticeable after the number of particles and the frictional properties have stabilised.

\blue{
Note that the most important metric to follow for optimisation convergence - especially in problems such as calibration where the objective, i.e. the discrepancy between simulation and experiment, is to be minimised as close as possible to a known value of zero (as opposed to minimisation where an objective is lowered to an unknown possible best value) - is the evolution of the error values. As seen in the last three panels of Figure \ref{fig:convergence}, during the last $\sim 6$ epochs all errors hovered around the minimum that was ultimately found. ACCES has a deliberately ``large'' default target tolerance of $0.1$ for the combined standard deviation, which follows the variation in final error; this was chosen in order to avoid the extensive hovering around the final minimum that optimisers are prone to - in many cases the conservative default parameters of general-purpose optimisers result in over half of the total iterations before termination to be spent around the optimum which was found much earlier \cite{rios2013derivative}. In all our numerical experiments, reducing the default target tolerance (relative to a normalised starting value of 1) below $0.1$ to $0.05$ and $0.01$ resulted in effectively the same calibrated parameters being found; to validate this in the present study, ACCES was restarted with the historical data of the 18 epochs shown in Figure \ref{fig:convergence} and an artificially-lowered target standard deviation, and manually stopped at epoch 26 - the same parameters were found to two significant digits, thus showing the diminishing returns of running optimisers to lower tolerances in calibration problems. Another nuanced point here is that the standard deviations shown in Figure \ref{fig:convergence} should be interpreted only relative to one another; this uncertainty is purely in the optimisation sense, and thus is distinct from experimental uncertainty, which naturally should be quantified with similarly experimental means. The uncertainty associated with numerical simulations can be broadly assessed using global sensitivity analysis tools, and, within a sufficiently small vicinity of the final optimum found - where variation in the objectives can reasonably well be approximated using the second-order multivariate polynomial of Response Surface Methodology or general Design of Experiments approaches - using classical statistical means.

Note that while the restitution does have the largest associated standard deviation - again, in an optimisation sense - the measurements are sensitive enough that a global, physical minimum can be determined; for a rigorous sensitivity analysis of the rotating drum with respect to restitution, we refer the reader to \cite{jenkins2024sensitivity}.
}

The final calibrated parameters found, corresponding to the lowest error achieved after 18 epochs (i.e. 144 script evaluations in total) are:

\begin{itemize}
    \item Sliding friction $\mu_s = 0.5945$.
    \item Rolling friction $\mu_r = 0.0433$.
    \item Restitution $\varepsilon = 0.6648$.
    \item Number of particles in 50 mL inside a GranuDrum $N_P = 33,458$.
\end{itemize}

\begin{figure*}[htbp]
    \centering
    \includegraphics[width=0.8\linewidth]{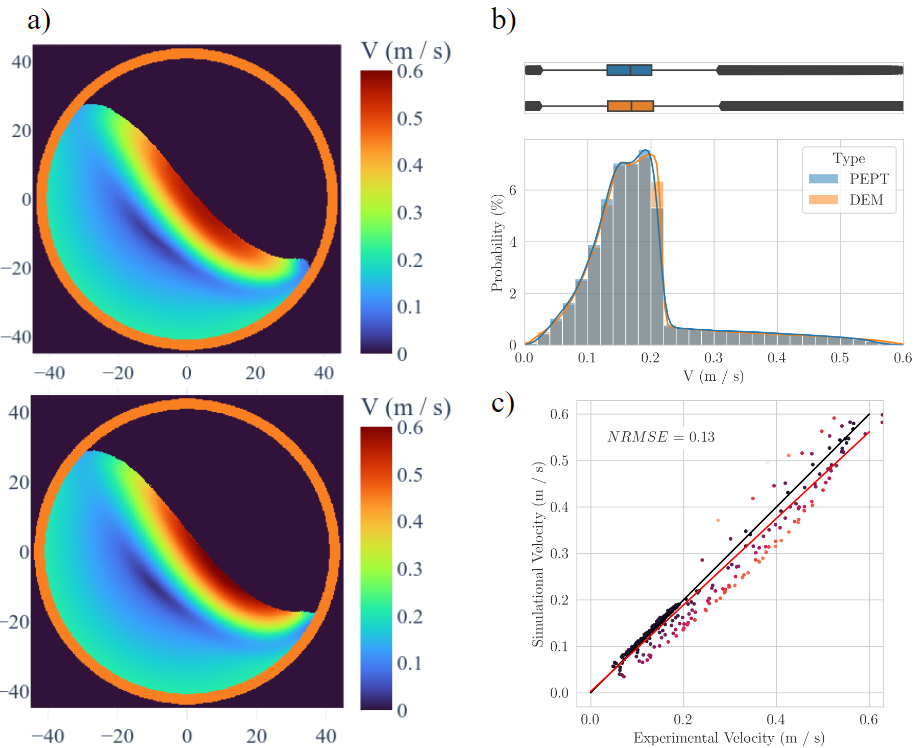}
    \caption{Velocity-based fields in a rotating drum at 50 RPM, from the calibrated DEM simulation and Positron Emission Particle Tracking (PEPT) experimental particle trajectories. Panel $a)$ shows a high-resolution velocity heatmap (computed using KonigCell \cite{Nicusan_KonigCell_Quantitative_Fast_2022}) from the PEPT data (top) and simulation (bottom). Panel $b)$ depicts histograms of the simulated and experimental velocities, along with box-plots including the Tukey fences, quartiles and medians. Panel $c)$ shows a parity plot comparing the velocities on a pixel-by-pixel basis from a $32 \times 32$ heatmap, along with the best-fit parity line (red) and the normalised root-mean-squared-error (NRMSE) of the predictions; \blue{the pixel-parity points are coloured by their distance from the identity line.}}
    \label{fig:granudrum_post}
\end{figure*}

\blue{It is worth noting that, assuming a suitable choice of calibration measurements (i.e.\ the problem is well-constrained), the values returned by ACCES are repeatable, with the precision of the agreement (i.e.\ the number of significant figures to which repeat values agree) determined by the final target standard deviation. In all of our numerical experiments -- using the specific methodology detailed in the paper -- where we varied both the population size, the initial random seed, and indeed the architecture on which the simulations were run, the same calibrated parameter values were found.}

Following the calibration at 10 and 50 RPM, the rotating drum can be further used for validation: i) a relatively easy test where the free surface area error is computed at a different rotation rate (as shown for 30 RPM in the bottom row of Figure \ref{fig:superimposed}); if it lies between the calibration rotation rates, it is expected to be readily reproduced; and ii) comparing other experimental measurements with simulated quantities which were not directly used for calibration. Velocity-based fields are particularly powerful as they define higher-order internal system dynamics which are difficult to reproduce without correct modelling of the physics, especially as calibration was conducted with very simple, binarised images. However, this requires that more extensive testing is conducted, with different imaging techniques; this paper details extensive validation in the next paragraphs with multiple quantities and distinct systems altogether, such that the ``base'' DEM calibration methodology can be applied to other powders by DEM practitioners without requiring time- and cost-intensive validation steps with other specialised imaging techniques -- rather, the simple binarised images of a rotating drum (commercially-available or lab-built) suffice. \blue{Note that the results in this paper focus on granular flows in dynamic systems; extending the effectiveness of the present DEM calibration to quasi-static, high-stress cases such as shear cells will form the basis of future studies.}

Figure \ref{fig:granudrum_post} shows velocity-based fields from the rotating drum, as imaged using Positron Emission Particle Tracking, which yields DEM-like data (see Section \ref{sec:pept}). Panel $a)$ shows good qualitative agreement between the velocity distributions of the experimental data (top) and the calibrated simulation (bottom) on a high-resolution 2D heatmap, with the main rotating drum regions of interest having high similarity: the medium-velocity solid-body moving region near the walls, the low-velocity centre, under the free-flowing surface, and the high-velocity free-flowing surface, especially its depth. The quantitative velocity probabilistic distributions shown in panel $b)$ show excellent agreement, with both distributions being almost indistinguishable between simulation and experiment.

Panel $c)$ of Figure \ref{fig:granudrum_post} depicts a more rigorous comparison between experimental PEPT data and calibrated DEM simulation data. An effective simulation -- that is, a comprehensive calibration methodology -- should accurately replicate the entire velocity distribution throughout the system volume. This can be evaluated by directly comparing each pixel from the experimental image -- each representing a velocity at a specific point in real space -- with its corresponding pixel in the DEM image -- each representing a velocity at a corresponding point in simulated space. If these values match within the experimental error margins across the entire system, it can be reasonably concluded that the simulation is accurately calibrated for the current setup. To \textit{quantitatively} gauge the level of agreement, a parity plot can be created, where each data point represents a comparison between a pair of matching simulated and experimental pixels. Ideally, for a perfectly accurate simulation, all points would align precisely with the identity line ($y=x$). Deviations from this line indicate lesser accuracy in the simulation. The accuracy of the calibration method can thus be quantitatively assessed by calculating the normalised root mean square error (NRMSE) of the identity line, where lower (higher) NRMSE values indicate better (poorer) performance. Here, the normal RMSE of the parity points relative to the identity line is normalised by the hypothetical RMSE that would result if there was a complete mismatch between the simulation and experiment -- i.e., if each pixel had a value of 0. In this case, an NRMSE value of $0.13$ value was achieved, indicating a very high level of agreement, and, as seen in Figure \ref{fig:granudrum_post}, the fitted parity line is very closely aligned to the identity line.

{It is worth emphasising here the importance of this result: the calibration process had no input regarding velocity, just static, binarised images and a measurement corresponding exclusively to the free-flowing surface shape, yet quantitatively predicts the velocity across the entire system!}

\subsection{Validation with Common Powder-Handling Systems}
\label{sec:validation}

Alongside the validation conducted in the previous section using higher-order quantities not directly used in the calibration methodology, we also implement the calibrated particle properties in DEM simulations of systems different to the one used for calibration, and compare them with PEPT-based experimental data. The success of this will imply that DEM is not merely a reduced-order model that cannot be trusted outside the ``training range'', but instead a high-fidelity simulation model with predictive capabilities -- and hence the particles can be ``taken out'' of the calibration system and ``placed into'' other equipment of interest. Importantly, we cover multiple orders of magnitude differences in scale, from 10 to 500 mL, as well as distinct operating regimes -- i.e. frictional in the rotating drum, frictional-collisional in the tumbling mixer, and collisional in the vibrofluidised bed.

\subsubsection{Tumbling Mixer}

\begin{figure*}[htbp]
    \centering
    \includegraphics[width=\linewidth]{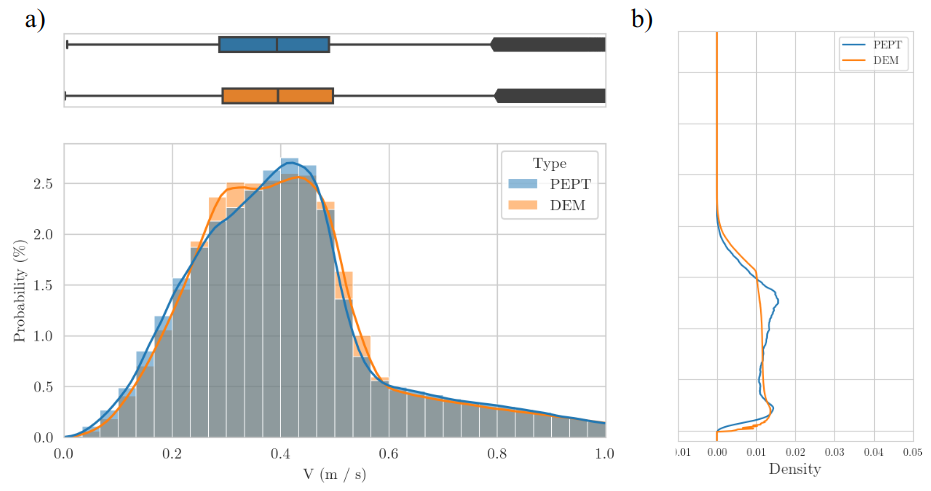}
    \caption{Quantities used for validating the ACCES-calibrated particle properties in a tumbling mixer: panel $a)$ shows the velocity histograms (bottom) and box plots (top) for the PEPT experiment and calibrated DEM simulation; panel $b)$ shows the vertical particle density, normalised by the integrated area; in this case the ``vertical'' axis corresponds to the mixer rotation axis, which lies at 60 degrees from the horizontal.}
    \label{fig:pascall_post}
\end{figure*}

Similar to the rotating drum analysis of the velocity distributions (see previous Section) between the DEM simulations and PEPT experiments, panel $a)$ of Figure \ref{fig:pascall_post} shows good agreement between the histograms; the box plots above them show very close matching of the quartiles and medians. 

Perhaps the key statistic of a tumbling mixer like the Pascall, where particles are lifted by the baffles, before sliding off them and free-falling back into the bulk, is its vertical occupancy. As for the velocities, panel $b)$ of Figure \ref{fig:pascall_post} shows good agreement between the vertical particle densities, with well-captured bottom regions (corresponding to solid bulk motion) and importantly the top, where both simulation and experiment show very similar curves tapering off, corresponding to matched particle sliding off the baffles. The slight kink shown in the PEPT curve in the upper-middle region is within experimental errors and may be due to lower tracer exploration of the bottom edges of the baffles, which are accounted for in the DEM data, as it includes all the particles simulated.

\subsubsection{Vibrofluidised Bed}
\label{sec:validation_resodyn}

\begin{figure*}[htbp]
    \centering
    \includegraphics[width=\linewidth]{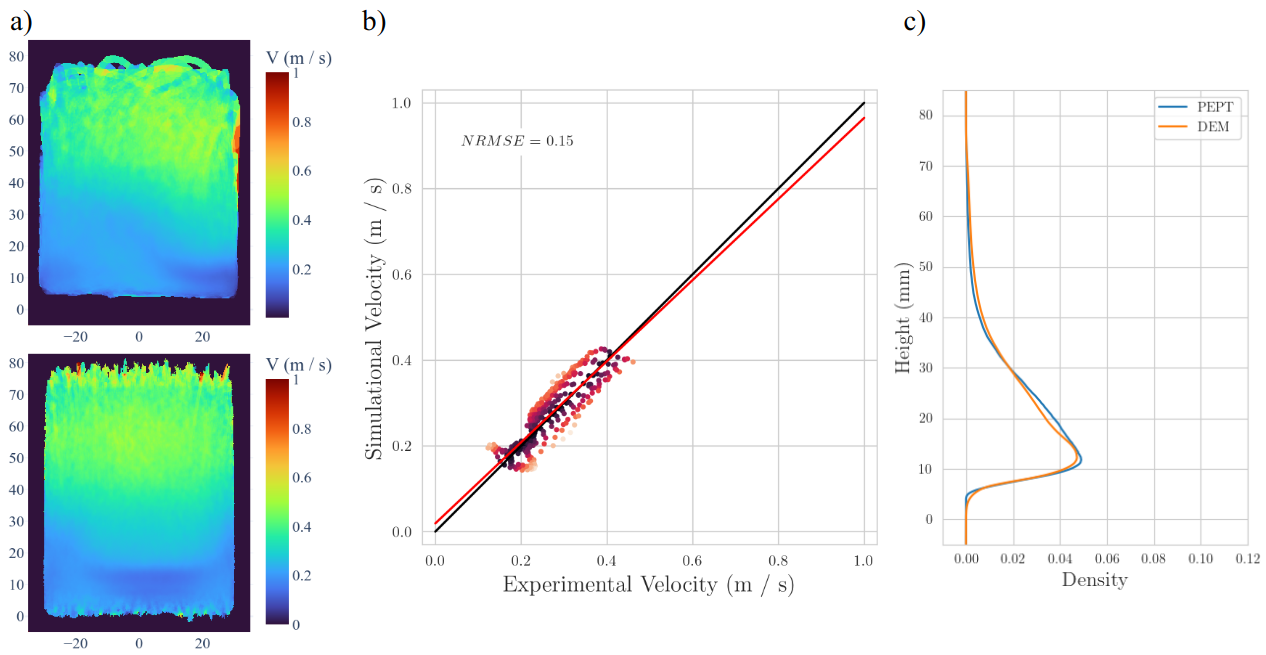}
    \caption{Quantities used for validating the ACCES-calibrated particle properties in a vibrofluidised bed: panel $a)$ shows high-resolution vertical velocity heatmaps (rasterisation done using KonigCell \cite{Nicusan_KonigCell_Quantitative_Fast_2022}) for the PEPT experiment (top) and the DEM simulation implementing the ACCES-calibrated particle properties (bottom). Panel $b)$ depicts a velocity-based parity plot computed pixel-by-pixel from a $32 \times 32$ pixel grid; the parity line of best fit is shown in red, while the identity line is black; \blue{the pixel-parity points are coloured by their distance from the identity line.}. Panel $c)$ outlines the vertical particle density profiles, normalised by their integrated area.}
    \label{fig:resodyn_post}
\end{figure*}

The Resodyn LabRAM II acoustic mixer is a lab-scale vibrofluidised bed shaking -- in this case -- 10 mL of powder at 62.44 Hz with 6.69 mm amplitude (full details in Section \ref{sec:experimental_systems}), thus operating distinctly in the collisional regime. As seen in the parameter uncertainty reduction in Figure \ref{fig:convergence}, restitution is the last particle property to be calibrated, as its effects are more subtle and only start becoming apparent once the bulk density and frictional properties are calibrated. The correctness -- i.e. physicality -- of the calibrated restitution can thus be validated by comparison with PEPT experimental data of this restitution-dominated system.

As shown in panel $a)$ of Figure \ref{fig:resodyn_post}, the high-resolution velocity heatmaps of the PEPT experiment (top) and DEM simulation (bottom) show good qualitative agreement, with similar relative heights of the dense bed at the bottom and high-velocity particle ``jumps'' above it. More quantitatively, the velocity parity plot shows close positioning of the pixel-by-pixel comparison to the identity line and a very low normalised RMSE value of 0.15 (full details on the parity-based method given in Section \ref{sec:converging}).

The most important defining characteristic of a restitution-dominated vertical vibrofluidised bed is its vertical occupancy, which is extremely sensitive to the precise value of the restitution coefficient \cite{windows2014inelasticity}. Panel $c)$ of Figure \ref{fig:resodyn_post} shows the vertical density profiles from the PEPT experiment (orange) and the DEM simulation implementing the ACCES-calibrated particle properties (blue) -- in this case, the two are virtually indistinguishable, thus comprehensively validating the correctness, and physicality of the ACCES-based calibration methodology of DEM simulations -- including for the difficult restitution parameter -- against simple dense particle flows in a rotating drum.

\section{Future Outlook and Conclusion}

In this paper we have introduced \textit{ACCES: Autonomous Calibration and Characterisation via Evolutionary Simulation}, a library for the automatic calibration of micro parameters of complex computer models against macro experimental measurements using evolutionary optimisation. It employs a unique metaprogramming-based software architecture for calibrating simulation \textit{scripts}, as opposed to traditional optimisation \textit{functions}, allowing ACCES to automatically parallelise the execution of arbitrary user simulation scripts on local workstations, supercomputing clusters, or custom setups, e.g. cloud computing instances. Thus, it can be non-invasively retrofitted to pre-existing simulations, with arbitrarily complex setup, execution and post-processing workflows. By modelling calibration as an optimisation problem, ACCES is agnostic to simulation engines and experimental calibration targets, and has been used to e.g. calibrate fluid simulations of plowshare mixers against positron emission particle tracking (PEPT) data using commercial CFD solvers \cite{hart2024autonomous}, and calibrate Monte-Carlo models of digitizers for gamma-ray detection \cite{herald2022autonomous}.

ACCES is published as part of the ``Coexist'' micro-macro calibration suite developed by the authors, written in pure, portable Python and continuously tested on Python versions 3.7 - 3.11, on Windows, MacOS and Ubuntu. The library has been registered as an open-source Python package on PyPI, released under the GNU General Public License v3.0; it is comprehensively documented with installation instructions, tutorials and a complete API manual, all hosted online on Read the Docs. It is the authors' hope that the accessibility of the library will enable experts and non-experts alike to use it for straightforward calibration and optimisation of computer models, beyond the subjects mentioned in this paper.

First, the importance and difficulty of computer model calibration were introduced (Section \ref{sec:introduction}). An overview of alternative calibration approaches is then given, focusing on common uses in conjunction with finite element analysis (FEA), computational fluid dynamics (CFD) and discrete element method (DEM) models across a wide variety of subjects; key advantages and disadvantages of calibration methods such as brute-force, design of experiments, neural networks, kriging / Gaussian Processes, Bayesian optimisation, and surrogate-based metamodelling in general, are highlighted. ACCES aims to solve their disadvantages by:

\begin{enumerate}
    \item Using optimiser-guided sampling targeted at globally-optimum calibrated parameters.
    \item Allowing efficient, parallel evaluation of samples.
    \item Using direct samples instead of surrogate models to avoid their approximation error altogether.
    \item No training needed, therefore not being sensitive to previously-seen systems.
    \item Minimising the number of function evaluations down to levels competitive with classical gradient-descent.
    \item Not having metaparameters to tune, ACCES can be applied directly to previously-unexplored problems.
    \item Implementing a novel, user-focused script-based ``simulation-in-the-loop'' software architecture.
\end{enumerate}

In Section \ref{sec:ACCES}, the framing of calibration as an optimisation problem was explained, followed by a detailed description of the novel ACCES software architecture, including abstract syntax tree (AST) manipulation of user simulation scripts, scheduler-agnostic parallel execution, the CMA-ES evolutionary optimiser and changes to its function for calibration, scalarising multiple objectives, saving deterministic traces of optimisation runs which can be resumed in case of node failure, and graceful fault-tolerance for potentially-crashing simulations.

ACCES was then applied to the infamously difficult problem of discrete element method (DEM) calibration for particulate systems, followed by extensive, rigorous validation (Section \ref{sec:methodology}). A novel calibration methodology was developed, simultaneously using two images of a rotating drum free-flowing surface, operating in two distinct regimes, to calibrate the sliding friction, rolling friction, restitution and number of particles in the drum. The latter effectively adds bulk density as an indirect calibration target, alongside the experimental-simulational superimposed free surface error areas (shown in Figure \ref{fig:twins}). The calibration was finished in under 24 hours on a consumer workstation with 8 cores, following 144 simulations in total, executed in epochs of 8 parallel evaluations at a time. We then validated the calibrated DEM particle properties in:

\begin{enumerate}
    \item The same drum, rotating at a different RPM than used for calibration, and including quantities that were not calibrated against: velocity heatmaps, histograms, and parity plots (Figure \ref{fig:granudrum_post}).
    \item A tumbling mixer with $10 \times$ larger scale than the drum, operating in a combined frictional-collisional regime, validating against the velocity histograms and vertical occupancies (Figure \ref{fig:pascall_post}).
    \item A vibrofluidised bed with $5 \times$ fewer particles than the drum, operating distinctly in the collisional regime, validating against velocity heatmaps, velocity parities and vertical occupancies (Figure \ref{fig:resodyn_post}).
\end{enumerate}

All systems were imaged using positron emission particle tracking (PEPT, illustrated in \ref{fig:pept}), yielding high-fidelity trajectories of radioactive, non-invasive tracers and allowing direct, 1-to-1 comparison between experiment and DEM simulation. Following the excellent agreement between the two, we have rigorously proven that:

\begin{enumerate}
    \item Using well-chosen, yet simple, readily-available measurements can be enough for comprehensive calibration of DEM simulations.
    \item Which can then reproduce the dynamics of systems at different scales and operating regimes to the calibration instrument, and, importantly, \textbf{can reproduce quantities that were not used in calibration}!
    \item DEM, with its ``inexact'' modelling of granular mechanics (see discussion in Section \ref{sec:alternative}), can quantitatively predict the internal dynamics of systems handling powders which were calibrated in a different instrument.
\end{enumerate}

\section*{Acknowledgements}
The computations described in this paper were performed using the University of Birmingham's BlueBEAR HPC service, which provides a High Performance Computing service to the University's research community. See http://www.birmingham.ac.uk/bear for more details.

The authors gratefully acknowledge the following funding: this work was supported by the EPSRC MAPP Future Manufacturing Hub in Manufacture using Advanced Powder Processes (2020); EPSRC Impact Acceleration Account (2021); EPSRC MAPP Future Manufacturing Hub in Manufacture using Advanced Powder Processes (2022).

\section*{CRediT Author Statement}
\textbf{Andrei-Leonard Nicușan}: conceptualisation, investigation, methodology, software, validation, visualisation, writing - original draft preparation. \textbf{Dominik Werner}: software, visualisation. \textbf{Jack Sykes}: investigation. \textbf{Jonathan Seville}: supervision, writing - review \& editing. \textbf{Tzany Kokalova-Wheldon}: resources. \textbf{Kit Windows-Yule}: conceptualisation, funding acquisition, project administration, resources, writing - review \& editing.

\section*{Declaration of competing interest}
The authors declare that they have no known competing financial interests or personal relationships that could have appeared to influence the work reported in this paper.

\section*{Supplementary Materials}
The particle size distributions, PEPT radiolabelled-tracer trajectories, ACCES calibration scripts and logs, LIGGGHTS DEM digital models of the GranuDrum rotating drum, Pascall tumbling mixer and ResoDyn acoustic mixer (including meshes, LIGGGHTS definition and launch scripts), and post-processing scripts are given in a Mendeley Data repository, DOI: 10.17632/ccpb9jd8j6.1.

\section*{Data Availability and Reproducibility Statement}
All data used in this paper, along with data-processing scripts and instructions to run and reproduce results are included in the Supplementary Materials. Specifically:

\begin{enumerate}
    \item The data for the particle size distributions (PSDs) shown in Figure \ref{fig:psd} are included in the ``ParticleSizeDistribution'' directory, including the raw CSV files from the Sympatec QicPic measurements and the Python scripts for PSD extraction and LIGGGHTS-script implementation.
    \item All experimental PEPT data is included in the ``ExperimentalPEPT'' directory, for the GranuDrum at 10 and 50 RPM, the ResoDyn acoustic mixer and Pascall tumbling mixer. This includes the raw PEPT-recorded radioactive tracer trajectories in the ``pept'' Python library format, saved as Python PICKLE files (e.g. ``gd\_p1\_10rpm\_trajs\_prep.pickle''), along with the Python processing scripts used for locating the tracers, computing higher-order quantities such as velocities, and post-processing the data to produce the PEPT plots in Figures \ref{fig:pept}, \ref{fig:granudrum_post}, \ref{fig:pascall_post}, and \ref{fig:resodyn_post}.
    \item Complete processing scripts and computation logs for the DEM simulations are included in the ``SimulationalDEM'' directory: \begin{enumerate}
        \item All ACCES-relevant scripts, data and logs are in ``SimulationalDEM/ACCES-Calibration''. The convergence plot shown in Figure \ref{fig:convergence} is created from the ACCES logs in the ``access\_seed42'' subdirectory, which, as detailed in Section \ref{sec:ACCES}, contains all results of the simulations executed, along with recorded standard and error outputs and timestamps. The LIGGGHTS scripts, ACCES scripts, post-processing scripts, launch configuration and measurements used for comparison are included; the best, calibrated parameters taken from the ACCES logs were used in the same scripts to produce the superimposed experimental-simulational images in Figure \ref{fig:superimposed}.
        \item The ``SimulationalDEM/GranuDrum50RPM'' subdirectory contains the calibrated simulation, including the LIGGGHTS script to run it, the configuration used on the BlueBEAR HPC, Python data-extraction code, higher-order quantity computation, and post-processing scripts for the occupancy, velocity heatmaps, histograms, and parity which form the underlying data of Figure \ref{fig:granudrum_post}.
        \item The ``SimulationalDEM/PascallTumblingMixer'' subdirectory contains the calibrated simulation of the digital model, including the LIGGGHTS script to run it, the configuration used on the BlueBEAR HPC, Python data-extraction code, higher-order quantity computation, and post-processing scripts for the vertical occupancy and velocity histogram which form the underlying data of Figure \ref{fig:pascall_post}.
        \item The ``SimulationalDEM/ResoDynAcousticMixer'' subdirectory contains the calibrated simulation of the digital model, including the LIGGGHTS script to run it, the configuration used on the BlueBEAR HPC, Python data-extraction code, higher-order quantity computation, and post-processing scripts for the vertical occupancy, velocity heatmaps and parity which form the underlying data of Figure \ref{fig:resodyn_post}.
    \end{enumerate}
\end{enumerate}

To make the \textit{use} of the methodology and codes developed in this paper accessible to a wider audience, we further include an online, interactive workshop on using LIGGGHTS and ACCES for simulation calibration and optimisation, available at the GitHub repository: https://github.com/uob-positron-imaging-centre/ACCES-GranuDrum-Calibration

\clearpage

\bibliographystyle{unsrt}
\bibliography{sample}


\end{document}